\documentclass{icmart}
\usepackage{epsfig}

\contact[smirnov@math.unige.ch]
{Universit\'e de Gen\`eve\\
Section de Math\'ematiques\\
2–4, rue du Li\`evre\\
Case postale 64\\
1211 Gen\`eve 4\\
Switzerland
}

\newtheorem*{lthm}{Loewner's theorem}
\newtheorem*{sch}{Schramm's principle}
\newtheorem*{pri}{Martingale principle}
\newtheorem*{propa}{(A) Conformal Invariance}
\newtheorem*{propb}{(B) Markov property}
\newtheorem*{propc}{(B') Conformal Markov property} 

\newtheorem{thm}{Theorem}

\newtheorem{prop}[thm]{Proposition}
\newtheorem{conj}[thm]{Conjecture}
\theoremstyle{definition}
\newtheorem{defi}[thm]{Definition}
\newtheorem{rem}[thm]{Remark}

\newcommand\unif{{\rightrightarrows}}
\def\wi{{w}}
\def\mesh{\epsilon}

\def\br#1{\left(#1\right)}
\def\sle#1{$\mathrm{SLE}\left(#1\right)$}
\def\slee{$\mathrm{SLE}$}

\def\brs#1{\left\{#1\right\}}
\def\abs#1{\left|#1\right|}

\def\Cal#1{{\cal#1}}

\newcommand{\ZZ}{{\mathbb Z}}

\newcommand{\Cp}{{\mathbb C_+}}
\newcommand{\EE}{{\mathbb E}}

\newcommand{\RR}{{\mathbb R}}
\newcommand{\IM}{{\mathrm{Im}}}

\def\Exp#1{{\EE\left(#1\right)}}

\title[Towards conformal invariance of 2D lattice models]{Towards conformal invariance\\ of 2D lattice models}

\author[Stanislav Smirnov]{Stanislav Smirnov}

\begin{document}
\begin{abstract}
Many 2D lattice models of physical phenomena
are conjectured to have
conformally invariant scaling limits:
percolation, Ising model, self-avoiding polymers,~\dots
This has led to numerous exact (but non-rigorous)
predictions of their scaling exponents and dimensions.
We will discuss how to prove the conformal 
invariance conjectures,
especially in relation to Schramm-Loewner Evolution.
\end{abstract}



\maketitle

\section{Introduction}\label{sec:intro}

For several 2D lattice models 
physicists were able to make a number
of spectacular predictions
(non-rigorous, but very convincing)
about exact values of various scaling exponents and dimensions.
Many methods were employed 
(Coulomb Gas, Conformal Field Theory, Quantum Gravity) 
with one underlying idea:
that the model at criticality 
has a continuum scaling limit (as mesh of the lattice goes to zero)
and the latter is conformally invariant.
Moreover, it is expected that there is 
only a one-parameter family of possible conformally invariant scaling limits,
so universality follows:
if the same model on different lattices (and sometimes at different temperatures)
has a conformally invariant scaling limit, it is necessarily the same.
Indeed, the two limits belong to the same one-parameter family,
and usually it directly follows that the corresponding parameter values coincide.

Recently mathematicians were able to offer  
different, perhaps better, and certainly more rigorous
understanding of those predictions, in many
cases providing proofs.
The point which is perhaps still less understood
both from mathematics and physics points of view
is why there exists a universal conformally invariant scaling limit.
However such behavior is supposed to be typical in 2D models at criticality:
Ising, percolation, self-avoiding polymers;
with universal conformally invariant curves arising as scaling limits
of the interfaces.

Until recently this was established only
for the scaling limit of the 2D random walk, the 2D Brownian motion.
This case is easier and somehow exceptional because of the Markov property.
Indeed, Brownian motion was originally constructed by Wiener \cite{Wiener},
and its conformal invariance 
(which holds in dimension 2 only)
was shown by Paul L\'evy \cite{Levy-book}
without appealing to random walk.
Note also that unlike interfaces
(which are often simple, or at most ``touch'' themselves), 
Brownian trajectory
has many ``transversal'' self-intersections.

For other lattice models even a rigorous formulation
of conformal invariance conjecture seemed elusive.
Considering percolation
(a model where vertices of a graph are declared open independently with equal probability $p$
-- see the discussion below)
at criticality as an example,
Robert Langlands, Philippe Pouliot and Yvan Saint-Aubin in \cite{Langlands-bams}
studied numerically crossing probabilities
(of events that there is an open crossing of a given rectangular shape).
Based on experiments they concluded
that crossing probabilities should have a universal
(independent of lattice) scaling limit, which is conformally invariant
(a conjecture they attributed to Michael Aizenman).
Thus the limit of crossing probability for a rectangular domain
should depend on its conformal modulus only.
Moreover an exact formula (\ref{eq:cardy}) using hypergeometric function 
was proposed by John Cardy in \cite{Cardy-92}
based on Conformal Field Theory arguments.
Later Lennart Carleson found that the formula has a particularly
nice form for equilateral triangles, see \cite{Smirnov}.
These developments got many researchers interested in the subject
and stimulated much of the subsequent progress.

Rick Kenyon \cite{Kenyon1,Kenyon2} established conformal invariance of many
observables related to dimer models (domino tilings),
in particular to uniform spanning tree and loop erased random walk,
but stopped short of constructing the limiting curves.

In \cite{Schramm-lerw}, Oded Schramm suggested to study the scaling limit
of a single interface and classified all possible
curves which can occur as conformally invariant scaling limits. 
Those turned out to be a universal one-parameter family
of \sle{\kappa} curves,
which are now called \emph{Schramm-Loewner Evolutions}.
The word ``evolution''
is used since the curves are constructed dynamically,
by running classical Loewner Evolution with
Brownian motion as a driving term.
We will discuss one possible setup,
chordal \sle{\kappa} with parameter $\kappa\in[0,\infty)$, 
which provides for each simply-connected domain $\Omega$ 
and boundary points $a$, $b$ a measure $\mu$ on curves from
$a$ to $b$ inside $\Omega$. The measures $\mu(\Omega,a,b)$
are conformally invariant, in particular they are all images
of one measure on a reference domain, say a half-plane $\Cp$.
An exact definition appears below. 

In \cite{Smirnov-cras,Smirnov}
the conformal invariance was established for critical percolation on triangular lattice.
Conformally invariant limit of the interface was identified with \sle6,
though its construction does not use \slee~ machinery.
See also Federico Camia and Charles Newman's paper \cite{Camia-Newman} for the 
details on subsequent construction of the full scaling limit.

In \cite{Lawler-Schramm-Werner-ust}
Greg Lawler, Oded Schramm and Wendelin Werner
have shown that a perimeter curve
of the uniform spanning tree converges to \sle8
(and the related loop erased random walk -- to \sle2)
on a general class of lattices.
Unlike the proof for percolation, theirs
utilizes \slee~ in a substantial way.
In \cite{Schramm-Sheffield} Oded Schramm and Scott Sheffield 
introduced a new model, Harmonic Explorer, 
where properties needed
for convergence to \sle4 
are built in.

Despite the results for percolation and uniform spanning tree,
the problem remained open for all other classical
(spin and random cluster)
2D models, including percolation on other lattices.
This was surprising given 
the abundance of the physics literature 
on conformal invariance. 
Perhaps most surprising was that the
problem of a conformally invariant scaling limit
remained open for the Ising model, 
since for the latter
there are many exact and often  rigorous results 
-- see the books \cite{Mccoy-Wu-book,Baxter-book}.

Recently we were able to work out the Ising case \cite{smirnov-fk1,smirnov-fk2,smirnov-fk3,smirnov-is}:
\begin{thm}
As lattice step goes to zero,
interfaces in Ising and Ising random cluster models 
on the square lattice at critical temperature
converge to~\sle3 and \sle{16/3} correspondingly.
\end{thm}
Computer simulations of these interfaces
(Figures~\ref{fig:ising},~\ref{fig:fkising})
as well as the definition of the Ising models
can be found below.
Similarly to mentioned experiments for percolation,
Robert Langlands, Marc Andr\'e Lewis and Yvan Saint-Aubin conducted
in \cite{Langlands-ising} numerical studies
of crossing probabilities for the Ising model at critical temperature.
A modification of the theorem above relating  interfaces to \slee's~
(with drifts) in
domains with five marked boundary points allows a rigorous
setup for establishing their conjectures.

The proof is based on showing that a
certain Fermionic lattice observable
(or rather two similar ones for spin and random cluster models)
is discrete analytic and solves a particular
covariant Riemann Boundary Value Problem.
Hence its limit is conformally covariant
and can be calculated exactly.
The statement is interesting in its own right,
and can be used to study spin correlations. 
The observable studied has more manifest physics meaning
than one in our percolation paper \cite{Smirnov}.

The methods lead to some progress in fairly
general families of random cluster and $O(n)$ models,
and not just on square lattices.
In particular, besides Ising cases, they seem 
to suggest new proofs for all other known cases
(i.e. site percolation on triangular lattice 
and  uniform spanning tree).

In this note we will discuss this proof and general
approach to scaling limits and conformal invariance of
interfaces in the \slee~context.
We will also state some of the open questions and
speculate on how one should approach other models.

We omit many aspects of this rich subject.
We do not discuss the general mathematical
theory of \slee~ curves or their connections
to physics, for which interested reader can consult 
expository works 
\cite{Bauer-Bernard,Cardy-expo,Nienhuis-sle,Werner-stflour}
and a book \cite{Lawler-book}.
We do not mention the question of how to deduce the values of scaling exponents for lattice models
with \slee~help once convergence is known. 
It was explored in some detail only for percolation \cite{Smirnov-Werner},
where convergence is known 
and  the required (difficult) estimates were already in place
thanks to Harry Kesten \cite{Kesten-scaling}.
We also  restrict ourselves to one interface, whereas
one can study the collection of all loops
(cf. exposition \cite{Werner-loops}), and many of our considerations
transfer to the loop soup observables.
Finally, 
there are many other open questions related to conformal invariance,
some of which are discussed in
Oded Schramm's paper \cite{Schramm-icm} in these proceedings.

\section{Lattice models}

We focus on two families of lattice models which have nice ``loop representations''.
Those families include or are closely related to most of
the ``important'' models, including
percolation, Ising, Potts, spherical (or $O(n)$),
Fortuin-Kasteleyn (or random cluster), 
self-avoiding random walk, and uniform spanning tree models.
For their interrelations and for the discussion of many other 
relevant models one can consult the books \cite{Baxter-book,Grimmett-book06,Madras-Slade,Mccoy-Wu-book}.
We also omit many references which can be found there.

There are various ways to understand the existence of the scaling limit and its conformal invariance.
One can ask for the full picture, which can be represented as a loop collection
(representing all cluster interfaces), random height function
(changing by $\pm1$ whenever we cross a loop),
or some other object. 
It however seems desirable to start with a simpler problem.

One can start with observables (like correlation functions,
crossing probabilities), for which it is easier to make sense of the limit:
there should exist a limit of a number sequence which 
is a conformal invariant.
Though a priori it might seem to be a weaker goal than constructing a full scaling limit,
there are indications that 
to obtain the full result
it might be sufficient to analyze just one observable.

We will discuss an intermediate goal to analyze the law of just one interface,
explain why working out just one observable would be sufficient,
and give details on how to find an observable with a conformally invariant limit.
To single out one interface, we consider a model on a simply connected domain 
with Dobrushin boundary conditions (which besides many
loop interfaces enforce existence of an interface joining two boundary points $a$ and $b$).
We omit the discussion of the full scaling limit,
as well as models on Riemann surfaces and with different boundary conditions.

\subsection{Percolation}
\begin{figure}[ht!]
\centerline{
\epsfxsize=12cm\epsfbox{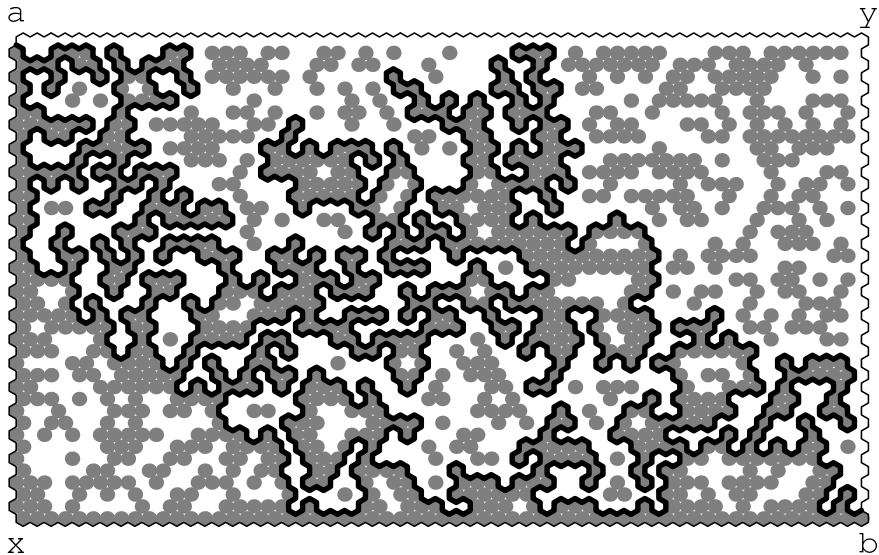}}
\caption{Critical site percolation on triangular lattice superimposed over a rectangle.
Every site is grey or white independently with equal probability $1/2$.
Dobrushin boundary conditions (grey on lower and left sides, white on upper and right sides)
produce an interface from the upper left corner $a$ to the lower right corner $b$.
The law of the interface converges to \sle6 when lattice step goes to zero,
while the rectangle is fixed. \label{fig:perc}}
\end{figure}

Perhaps the simplest  model (to state) is Bernoulli percolation on the triangular lattice.
Vertices are declared open or closed 
(grey or white in Figure~\ref{fig:perc})
independently with probabilities $p$ and $(1-p)$ correspondingly.
The critical value is $p=p_c=1/2$ -- see \cite{Kesten-book,Grimmett-book99},
in which case all colorings are equally probable.

Then each configuration can be represented by a collection of interfaces
-- loops which go along the edges of the dual hexagonal lattice and separate
open and closed vertices. 

We want to distinguish one particular interface, and to this effect
we introduce Dobrushin boundary conditions:
we take two boundary points $a$ and $b$ in a simply connected
$\Omega$ (or rather its lattice approximation),
asking the counterclockwise arc $ab$ to be grey
and the counterclockwise arc $ba$ to be white.
This enforces existence of a single non-loop interface
which runs from $a$ to $b$.
The ``loop gas'' formulation of our model  is that we consider all collections 
of disjoint loops plus a curve from $a$ to $b$
on hexagonal lattice with equal probability.

For each value of the lattice step $\mesh>0$ we approximate a given domain $\Omega$ 
by a lattice domain, which leads to a random interface,
that is a probability measure $\mu_\mesh$
on curves (broken lines) running from $a$ to $b$.
The question is whether there is a limit measure $\mu=\mu(\Omega,a,b)$
on curves and whether it is conformally invariant.
To make sense of the limit we consider the curves with uniform topology
 generated by parameterizations
(with distance between $\gamma_1$ and $\gamma_2$
being $\inf\|f_1-f_2\|_\infty$
where the infimum is taken over all
parameterizations $f_1$, $f_2$ of $\gamma_1$, $\gamma_2$), 
and
ask for weak-$*$ convergence of the measures $\mu_\mesh$.

\subsection{$O(n)$ and loop models}\label{sec:on}

Percolation turns out to be a particular case of the \emph{loop gas}
model which is closely related (via high-temperature expansion) 
to $O(n)$ 
(spherical) model. 
We consider configurations of non-intersecting
simple loops and a curve running 
from $a$ to $b$ on \emph{hexagonal lattice} inside domain $\Omega$
as for percolation in Figure~\ref{fig:perc}.
But instead of asking all configurations to be equally likely,
we introduce two parameters:
loop-weight $n\ge0$ and edge-weight $x>0$, and ask
that probability of a configuration is proportional to
$$
n^{{\mathrm{\#~loops}}}\,x^{{\mathrm{ length~of~loops}}}~.
$$
The vertices not visited by loops are called monomers.
Instead of weighting edges by $x$ one can equivalently weight monomers by $1/x$.

We are interested in the range $n\in[0,2]$
(after certain modifications $n\in[-2,2]$ would work),
where conformal invariance is expected
(other values of $n$ have different behavior).
It turns out that there is a critical value $x_c(n)$,
such that the model
exhibits one critical behavior at $x_c(n)$
and another on the interval $(x_c(n),+\infty)$,
corresponding to ``dilute'' and ``dense'' phases
(when in the limit the loops are simple and non-simple correspondingly).

Bernard Nienhuis \cite{N1,N2} proposed the following conjecture,
supported by physics arguments:
\begin{conj}\label{conj:nienhuis}
The critical value is given by
$$x_{c}(n)=\frac1{\sqrt{2+\sqrt{2-n}}}~.$$
\end{conj}
\noindent
Note that though for all $x\in(x_c(n),\infty)$ the critical behavior (and the scaling limit)
are conjecturally the same, the related value
$\tilde x_{c}(n)=1/{\sqrt{2-\sqrt{2-n}}}$ turns out to be distinguished 
in some ways.

The criticality was rigorously established for $n=1$ only,
but we still may discuss the scaling limits at those values of $x$.
It is widely believed that at the critical values the model 
has a conformally invariant scaling limit.
Moreover,  the corresponding criticalities under renormalization
are supposed to be unstable and stable correspondingly,
so for $x=x_c$
there should be one conformally invariant scaling limit,
whereas for the interval
$x\in(x_c,\infty)$ another, corresponding to $\tilde x_c$.
The scaling limit for low temperatures $x\in(0,x_c)$,
a straight segment,
is not conformally invariant.

\begin{figure}[ht!]
\centering{\epsfxsize=10cm\epsfbox{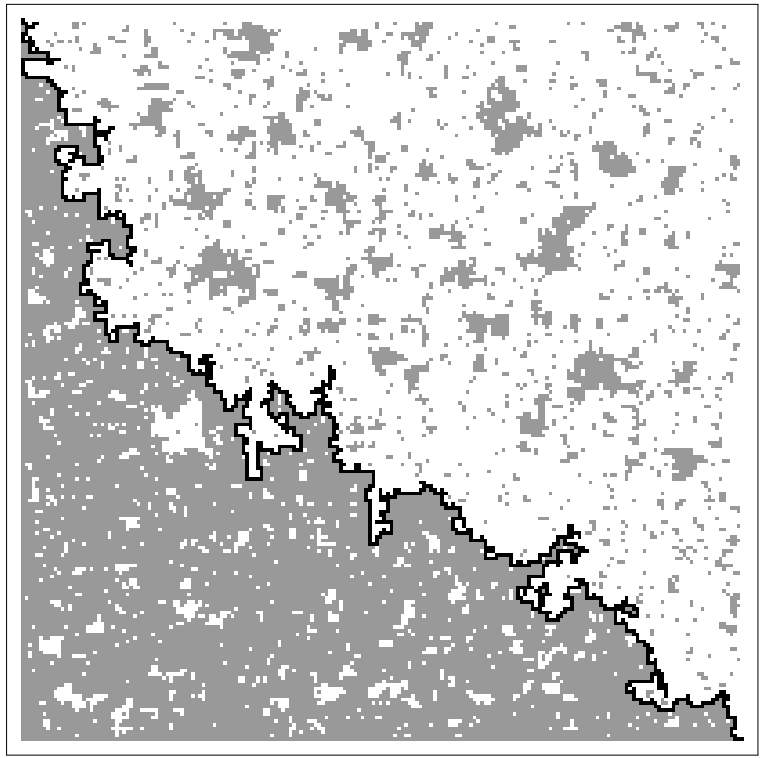}}
\caption{The Ising model at critical temperature on a square.
White and grey sites represent $\pm1$ spins.
Dobrushin boundary conditions (grey on lower and left sides, white on upper and right sides)
produce, besides loop interfaces, an interface from the upper left to the lower right corner,
pictured in black.
When lattice step goes to zero, the law of the interface converges to \sle3,
which is a conformally invariant  random curve,
almost surely simple and of Hausdorff dimension $11/8$.
\label{fig:ising}}
\end{figure}

Plugging in $n=1$ we obtain weight
$$
x^{{\mathrm{ length~of~loops}}}~.
$$
Assigning the spins $\pm1$ 
(represented by grey and white colors in Figure~\ref{fig:perc})
to sites of triangular lattice,
we rewrite the weight as
\begin{equation}\label{eq:is}
x^{\mathrm{\#~pairs~of~neighbors~of~opposite~spins}}~,
\end{equation}
obtaining the  Ising model (where the usual
parameterization is $\exp(-2\beta)=x$).
The critical value is known to be $\beta_c=\log3/4$,
so one gets the Ising model at critical  temperature for $n=1$, $x=1/\sqrt{3}$.
A computer simulation of the Ising model on the square lattice
at critical temperature, when
the probability of configuration is proportional to
(\ref{eq:is}),
is shown in Figure~\ref{fig:ising}.

For $n=1$, $x=1$ we obtain critical site percolation on triangular lattice.
Taking $n=0$ (which amounts to considering configurations with no loops,
just a curve running from $a$ to $b$),
one obtains for $x_c=1/\sqrt{2+\sqrt{2}}$ a version
of the self-avoiding random walk.

The following conjecture
(see e.g. \cite{Nienhuis-sle})
is a direct consequence
of physics predictions and \slee~ calculations:

\begin{conj}\label{conj:on}
For $n\in[0,2]$ and $x=x_c(n)$, 
as lattice step goes to zero,
the law of the interface converges to
Schramm-Loewner Evolution with
$$\kappa=4\pi/(2\pi-\arccos(-n/2))~.$$
For $n\in[0,2]$ and $x\in(x_c,\infty)$
(in particular for $x=\tilde x_c$), 
as lattice step goes to zero,
the law of the interface converges to
Schramm-Loewner Evolution with
$$\kappa=4\pi/\arccos(-n/2)~.$$
\end{conj}
Note that to address this question one does not need to prove
that the Nienhuis temperature is indeed critical
(Conjecture~\ref{conj:nienhuis}).

We discussed loops on the hexagonal lattice,
since it is a trivalent graph and so at most one
interface can pass through a vertex.
One can engage in similar considerations
on the square lattice with special regard to a possibility of two interfaces passing through the same vertex,
in which case they can be split into loops in two different ways
(with different configuration weights).
In the case of Ising ($n=1$) this
poses less of a problem, since number of loops is not important.
For $n=1$ and $x=1$ we get percolation model with $p=1/2$,
but for a general lattice this $p$ need not be critical,
so e.g. critical site percolation on the square lattice does not fit directly into this framework.

\subsection{Fortuin-Kasteleyn random cluster models}
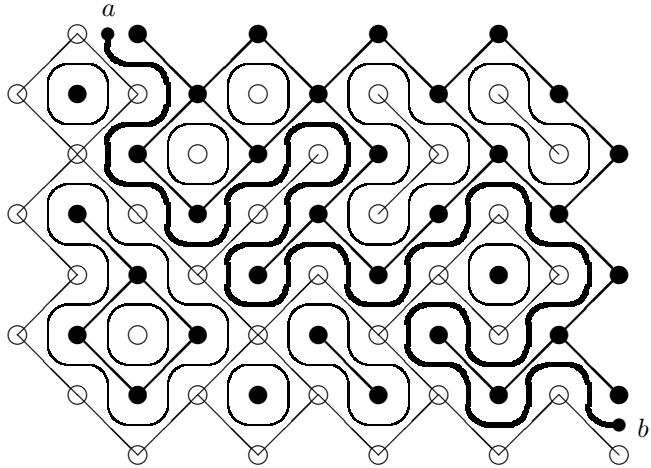
\begin{figure}\centering
{\def\bs{\circle*{3}}\def\ws{\circle{3}}
\def\ne{\qbezier(-0.,3)(0.5,0.5)(3,0)}
\def\nw{\qbezier(-0.,3)(-0.5,0.5)(-3,0)}
\def\se{\qbezier(-0.,-3)(0.5,-0.5)(3,0)}
\def\sw{\qbezier(-0.,-3)(-0.5,-0.5)(-3,0)}
\def\bne{\qbezier(-0.3,3)(0.2,0.5)(3,-0)}
\def\bnw{\qbezier(-0.3,3)(-0.8,0.5)(-3,0)}
\def\bse{\qbezier(-0.3,-3)(0.2,-0.5)(3,0)}
\def\bsw{\qbezier(-0.3,-3)(-0.8,-0.5)(-3,0)}
\def\uu{{\thicklines\line(1,1){10}}}
\def\dd{{\thicklines\line(1,-1){10}}}
\def\uw{\line(1,1){10}}
\def\dw{\line(1,-1){10}}
\unitlength=0.8mm
\begin{picture}(115,85)(-10,-10)
\put(-5,15)\uw\put(-5,35)\uw\put(-5,55)\uw\put(05,45)\uw\put(15,-5)\uw
\put(25,05)\uw\put(25,25)\uw\put(35,-5)\uw\put(35,15)\uw\put(35,35)\uw\put(55,-5)\uw
\put(55,15)\uw\put(55,35)\uw\put(65,25)\uw\put(75,-5)\uw\put(75,15)\uw\put(85,05)\dw
\put(-5,15)\dw\put(-5,35)\dw\put(-5,55)\dw\put(05,05)\dw\put(05,45)\dw\put(05,65)\dw
\put(15,35)\dw\put(25,05)\dw\put(25,25)\dw\put(35,15)\dw\put(45,05)\dw\put(45,25)\dw
\put(55,15)\dw\put(55,55)\dw\put(65,25)\dw\put(65,05)\dw\put(75,55)\dw\put(75,35)\dw
\put(15,45)\uu\put(05,15)\uu\put(15,05)\uu\put(25,35)\uu\put(25,55)\uu\put(35,45)\uu
\put(45,35)\uu\put(45,55)\uu\put(55,25)\uu\put(65,35)\uu\put(65,55)\uu\put(75,05)\uu
\put(85,15)\uu\put(85,35)\uu\put(35,25)\uu
\put(05,15)\dd\put(05,35)\dd\put(15,25)\dd\put(15,45)\dd\put(25,55)\dd\put(35,65)\dd
\put(45,15)\dd\put(45,35)\dd\put(45,55)\dd\put(55,65)\dd\put(65,15)\dd\put(85,15)\dd
\put(65,55)\dd\put(75,45)\dd\put(75,65)\dd\put(85,55)\dd\put(85,35)\dd\put(15,65)\dd
{\linethickness{1.5pt}
\put(10,65){\line(0,-1){2}}\put(10,47){\line(0,-1){4}}\put(20,37){\line(0,-1){4}}\put(20,57){\line(0,-1){4}}
\put(30,37){\line(0,-1){4}}\put(30,27){\line(0,-1){4}}\put(40,47){\line(0,-1){4}}\put(40,37){\line(0,-1){4}}
\put(40,27){\line(0,-1){4}}\put(50,47){\line(0,-1){4}}\put(50,27){\line(0,-1){4}}\put(60,27){\line(0,-1){4}}
\put(60,17){\line(0,-1){4}}\put(70,37){\line(0,-1){4}}\put(70,17){\line(0,-1){4}}\put(70,7){\line(0,-1){4}}
\put(80,37){\line(0,-1){4}}\put(80,17){\line(0,-1){4}}\put(80,7){\line(0,-1){4}}\put(90,27){\line(0,-1){4}}
\put(90,07){\line(0,-1){4}}\put(73,0){\line(1,0){4}}\put(63,10){\line(1,0){4}}\put(83,10){\line(1,0){4}}
\put(33,20){\line(1,0){4}}\put(53,20){\line(1,0){4}}\put(63,20){\line(1,0){4}}\put(83,20){\line(1,0){4}}
\put(93,00){\line(1,0){2}}\put(23,30){\line(1,0){4}}\put(33,30){\line(1,0){4}}\put(63,30){\line(1,0){4}}
\put(83,30){\line(1,0){4}}\put(13,40){\line(1,0){4}}\put(33,40){\line(1,0){4}}\put(43,40){\line(1,0){4}}
\put(73,40){\line(1,0){4}}\put(13,50){\line(1,0){4}}\put(43,50){\line(1,0){4}}\put(13,60){\line(1,0){4}}
\put(43,30){\line(1,0){4}}\put(73,10){\line(1,0){4}}
\put(10,60)\bne\put(10,40)\bne\put(20,30)\bne\put(30,20)\bne\put(50,20)\bne\put(60,10)\bne
\put(70,00)\bne\put(70,10)\bne\put(80,30)\bne
\put(20,50)\bnw\put(30,30)\bnw\put(40,20)\bnw\put(40,40)\bnw\put(50,40)\bnw\put(60,20)\bnw
\put(70,30)\bnw\put(80,00)\bnw\put(90,10)\bsw\put(90,20)\bnw\put(80,10)\bnw\put(40,30)\bnw
\put(20,60)\bsw\put(20,40)\bsw\put(50,30)\bsw\put(50,50)\bsw\put(70,10)\bsw\put(70,20)\bsw
\put(80,40)\bsw\put(90,30)\bsw
\put(10,50)\bse\put(30,30)\bse\put(30,40)\bse\put(40,30)\bse\put(40,40)\bse\put(40,50)\bse
\put(60,20)\bse\put(60,30)\bse\put(70,40)\bse\put(80,10)\bse\put(80,20)\bse\put(90,0)\bne
}
\put(10,00)\ne\put(30,00)\ne\put(50,00)\ne\put(00,10)\ne\put(40,10)\ne\put(50,10)\ne
\put(20,20)\ne\put(70,20)\ne\put(00,30)\ne\put(10,30)\ne\put(50,30)\ne\put(20,40)\ne
\put(80,40)\ne\put(00,50)\ne\put(30,50)\ne\put(50,50)\ne\put(60,50)\ne\put(70,50)\ne
\put(80,50)\ne\put(10,10)\ne
\put(20,00)\nw\put(40,00)\nw\put(60,00)\nw\put(20,10)\nw\put(30,10)\nw\put(10,20)\nw
\put(80,20)\nw\put(60,30)\nw\put(30,40)\nw\put(70,40)\nw\put(90,40)\nw\put(10,50)\nw
\put(40,50)\nw\put(60,40)\nw
\put(10,10)\sw\put(40,10)\sw\put(50,10)\sw\put(60,10)\sw\put(20,20)\sw\put(30,20)\sw
\put(50,20)\sw\put(10,30)\sw\put(20,30)\sw\put(80,30)\sw\put(10,40)\sw\put(60,50)\sw
\put(70,50)\sw\put(80,50)\sw\put(90,50)\sw\put(10,60)\sw\put(40,60)\sw\put(60,60)\sw
\put(80,60)\sw\put(30,50)\sw
\put(20,10)\se\put(30,10)\se\put(00,20)\se\put(10,20)\se\put(40,20)\se\put(70,30)\se
\put(00,40)\se\put(50,40)\se\put(20,50)\se\put(00,60)\se\put(30,60)\se\put(60,40)\se
\put(50,60)\se\put(70,60)\se
\put(00,17){\line(0,-1){4}}\put(00,37){\line(0,-1){4}}\put(00,57){\line(0,-1){4}}\put(10,07){\line(0,-1){4}}
\put(10,17){\line(0,-1){4}}\put(10,27){\line(0,-1){4}}\put(10,37){\line(0,-1){4}}\put(10,47){\line(0,-1){4}}
\put(20,07){\line(0,-1){4}}\put(20,17){\line(0,-1){4}}\put(20,27){\line(0,-1){4}}\put(20,47){\line(0,-1){4}}
\put(30,07){\line(0,-1){4}}\put(30,17){\line(0,-1){4}}\put(30,47){\line(0,-1){4}}\put(30,57){\line(0,-1){4}}
\put(40,07){\line(0,-1){4}}\put(40,17){\line(0,-1){4}}\put(40,57){\line(0,-1){4}}\put(50,07){\line(0,-1){4}}
\put(50,17){\line(0,-1){4}}\put(50,37){\line(0,-1){4}}\put(50,57){\line(0,-1){4}}\put(60,07){\line(0,-1){4}}
\put(60,37){\line(0,-1){4}}\put(60,57){\line(0,-1){4}}\put(60,47){\line(0,-1){4}}\put(70,27){\line(0,-1){4}}
\put(70,47){\line(0,-1){4}}\put(70,57){\line(0,-1){4}}\put(80,27){\line(0,-1){4}}\put(80,47){\line(0,-1){4}}
\put(80,57){\line(0,-1){4}}\put(90,47){\line(0,-1){4}}\put(10,57){\line(0,-1){4}}
\put(13,00){\line(1,0){4}}\put(33,00){\line(1,0){4}}\put(53,00){\line(1,0){4}}\put(03,10){\line(1,0){4}}
\put(13,10){\line(1,0){4}}\put(23,10){\line(1,0){4}}\put(33,10){\line(1,0){4}}\put(43,10){\line(1,0){4}}
\put(53,10){\line(1,0){4}}\put(03,20){\line(1,0){4}}\put(13,20){\line(1,0){4}}\put(23,20){\line(1,0){4}}
\put(43,20){\line(1,0){4}}\put(03,30){\line(1,0){4}}\put(13,30){\line(1,0){4}}\put(53,30){\line(1,0){4}}
\put(53,40){\line(1,0){4}}\put(63,40){\line(1,0){4}}\put(83,40){\line(1,0){4}}\put(03,50){\line(1,0){4}}
\put(03,40){\line(1,0){4}}\put(23,40){\line(1,0){4}}\put(73,30){\line(1,0){4}}\put(73,20){\line(1,0){4}}
\put(23,50){\line(1,0){4}}\put(33,50){\line(1,0){4}}\put(53,50){\line(1,0){4}}\put(63,50){\line(1,0){4}}
\put(73,50){\line(1,0){4}}\put(83,50){\line(1,0){4}}\put(03,60){\line(1,0){4}}\put(33,60){\line(1,0){4}}
\put(53,60){\line(1,0){4}}\put(73,60){\line(1,0){4}}
\put(09,68){$a$}\put(98,-2){$b$}
\put(10,65){\circle*{2}}
\put(95,00){\circle*{2}}
\put(15,05)\bs\put(35,05)\bs\put(55,05)\bs\put(75,05)\bs
\put(15,25)\bs\put(35,25)\bs\put(55,25)\bs
\put(75,25)\bs\put(95,25)\bs\put(95,05)\bs
\put(15,45)\bs\put(35,45)\bs\put(55,45)\bs
\put(75,45)\bs\put(95,45)\bs
\put(15,65)\bs\put(35,65)\bs\put(55,65)\bs\put(75,65)\bs
\put(05,15)\bs\put(25,15)\bs\put(45,15)\bs\put(65,15)\bs\put(85,15)\bs
\put(05,35)\bs\put(25,35)\bs\put(45,35)\bs\put(65,35)\bs\put(85,35)\bs
\put(05,55)\bs\put(25,55)\bs\put(45,55)\bs\put(65,55)\bs\put(85,55)\bs
\put(05,05)\ws\put(25,05)\ws\put(45,05)\ws\put(65,05)\ws\put(85,05)\ws
\put(05,25)\ws\put(25,25)\ws\put(45,25)\ws\put(65,25)\ws\put(85,25)\ws
\put(05,45)\ws\put(25,45)\ws\put(45,45)\ws\put(65,45)\ws\put(85,45)\ws
\put(05,65)\ws
\put(-5,15)\ws\put(15,15)\ws\put(35,15)\ws\put(55,15)\ws\put(75,15)\ws
\put(-5,35)\ws\put(15,35)\ws\put(35,35)\ws\put(55,35)\ws\put(75,35)\ws
\put(-5,55)\ws\put(15,55)\ws\put(35,55)\ws\put(55,55)\ws\put(75,55)\ws
\put(95,-5)\ws
\put(15,-5)\ws\put(35,-5)\ws\put(55,-5)\ws\put(75,-5)\ws
\end{picture}}
\caption{
Loop representation
of the random cluster model.
The sites of the original  lattice are colored in black,
while the sites of the dual lattice are colored in white.
Clusters and dual clusters and loops separating them are pictured.
Under Dobrushin boundary conditions
besides a number of loops there is an interface running from $a$ to $b$, which is drawn in bold.
Weight of the configuration is proportional
to $(\sqrt{q})^{\mathrm{ \#~loops}}$.}
\label{fig:loops}
\end{figure}

Another interesting class is Fortuin-Kasteleyn models,
which are random cluster representations of $q$-state Potts model.
The random cluster measure on a graph (a piece of the \emph{square lattice} in our case)
is a probability measure on edge configurations (each edge is declared either open or closed),
such that the probability of a configuration is proportional to
$$p^{\mathrm{ \#~open~edges}}~(1-p)^{\mathrm{ \#~closed~edges}}~q^{\mathrm{ \#~clusters}}~,$$
where clusters are maximal subgraphs connected by open edges.
The two parameters are edge-weight $p\in[0,1]$ and cluster-weight $q\in(0,\infty)$,
with $q\in[0,4]$ being interesting in our framework
(similarly to the previous model, $q>4$
exhibits different behavior).
For a square lattice (or in general any planar graph) to every configuration
one can prescribe a cluster configuration on the dual graph, such that
every open edge is intersected by a dual closed edge and vice versa.
See Figure~\ref{fig:loops} for a picture of two dual configurations
with respective open edges.
It turns out that the probability of a dual configuration becomes proportional to
$$p_*^{\mathrm{\#~dual~open~edges}}~(1-p_*)^{\mathrm{\#~dual~closed~edges}}~q^{\mathrm{\#~dual~clusters}}~,$$
with the dual to $p$ value $p_*=p_*(p)$ satisfying $p_*/(1-p_*)=q(1-p)/p$.
For $p=p_{\mathrm{sd}}:=\sqrt{q}/(\sqrt{q}+1)$ the dual value coincides with the original one:
one gets $p_{\mathrm{sd}}=(p_{\mathrm{sd}})_*$ and so the model is self-dual.
It is conjectured that this is also the critical value of $p$,
which was only proved for $q=1$ (percolation), $q=2$ (Ising) and $q>25.72$.

Again we introduce Dobrushin boundary conditions:
wired on the counterclockwise arc $ab$
(meaning that all edges along the arc are open)
and dual-wired on the counterclockwise arc $ba$
(meaning that all dual edges along the arc are open,
or equivalently all primal edges orthogonal to the arc are closed)
- see Figure~\ref{fig:loops}.
Then 
there is a unique interface running from $a$ to $b$, 
which separates cluster containing the arc $ab$ from the dual cluster containing the arc $ba$.

We will work with the loop representation, which is similar to that in \ref{sec:on}.
The cluster configurations can be represented as Hamiltonian (i.e. including all edges)
non-intersecting (more precisely, there are no ``transversal'' intersections) loop configurations on the medial lattice.
The latter is a square lattice which has edge centers of the original lattice as vertices.
The loops represent interfaces between
cluster and dual clusters and turn by $\pm\frac\pi2$
at every vertex -- see Figure~\ref{fig:loops}.
It is well-known that probability of a configuration
is proportional to 
$$\br{\frac{p}{1-p}\frac1{\sqrt{q}}}^{\mathrm{ \#~open~edges}}\cdot\br{\sqrt{q}}^{\mathrm{ \#~loops}}~,$$
which for the self-dual value $p=p_{sd}$ simplifies to
\begin{equation}\label{eq:fk}
\br{\sqrt{q}}^{\mathrm{ \#~loops}}~.
\end{equation}
Dobrushin boundary conditions amount to introducing
two vertices with odd number of edges:
a source $a$ and a sink $b$,
which enforces a curve running form $a$ to $b$
(besides loops) -- see Figure~\ref{fig:loops}
for a typical configuration.

\begin{conj}\label{conj:fk}
For all $q\in[0,4]$, as the lattice step goes to zero,
the law of the interface converges to Schramm-Loewner Evolution with
$\kappa=4\pi/\arccos(-\sqrt{q}/2)$.
\end{conj}

Conjecture was proved by Greg Lawler, Oded Schramm and Wendelin Werner 
\cite{Lawler-Schramm-Werner-ust}
for the case of $q=0$,
when they showed that the perimeter curve of the uniform spanning tree
converges to \sle8.
Note that with Dobrushin boundary conditions
loop representation still makes sense for $q=0$.
In fact, the formula (\ref{eq:fk})
means that we restrict ourselves to configurations with no loops, just a curve
running from $a$ to $b$ 
(which then necessarily passes through all the edges),
and all configurations are equally probable.

\begin{figure}[ht!]
\centering{\epsfxsize=10cm\epsfbox{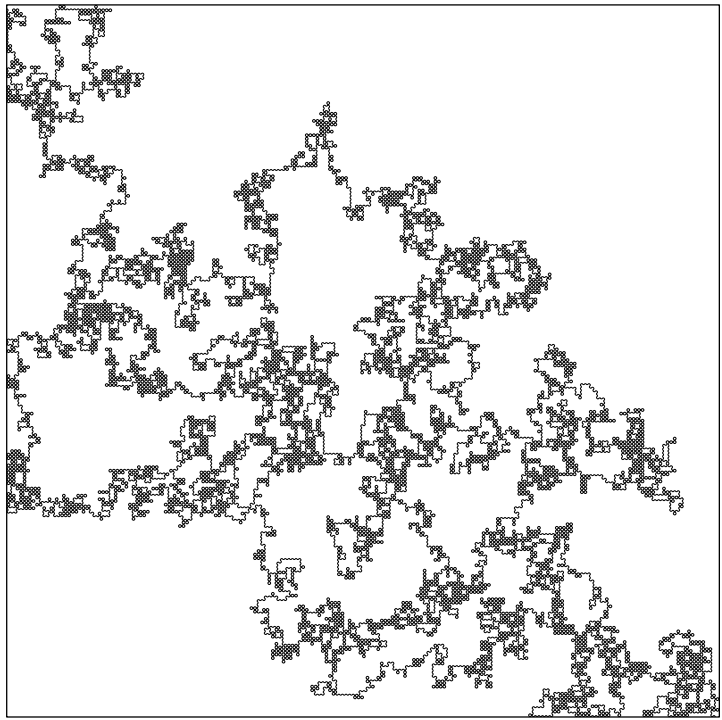}}
\caption{Interface in the random cluster Ising model 
at critical temperature with
Dobrushin boundary conditions (loops not pictured).
The law converges to \sle{16/3} when mesh goes to zero,
so in the limit it has Hausdorff dimension $5/3$ 
and touches itself almost surely.
The random cluster is obtained by deleting some bonds
from the spin cluster, so the interfaces are naturally different.
Indeed, they  converge to different \slee's
and have different dimensions.
However they are related: conjecturally,
the outer boundary of the 
(non-simple) pictured curve 
and the (simple) spin interface in Figure~\ref{fig:ising}
have the same limit after appropriate conditioning.
 \label{fig:fkising}}
\end{figure}

Below we will outline our proof
\cite{smirnov-fk1,smirnov-fk2}
that for the Ising parameter $q=2$
the interface converges to \sle{16/3},
see Figure~\ref{fig:fkising}.
It almost directly translates into a proof
that the interface of the spin cluster
for the Ising model on the square lattice at the critical temperature
(which can be rewritten as the loop model in \ref{sec:on} for $n=1$, only on the square lattice)
converges to \sle3, as shown on Figure~\ref{fig:ising}.
It seems likely that it will work in the $n=1$ case for the loop
model on hexagonal lattice described above,
providing convergence to \sle3 for $x=x_c$ and (a new proof of)
convergence to \sle6 for $x=\tilde x_c$
(and possibly for all $x>x_c$).

Summing it up, Conjecture~\ref{conj:on}
was proved earlier for $n=1$, $x=\tilde x_c$, see \cite{Smirnov,Smirnov-cras},
whereas Conjecture~\ref{conj:fk} was established
for $q=0$, see \cite{Lawler-Schramm-Werner-ust}.
We outline a technique, which seems to
prove conformal invariance in two new cases,
and provide new proofs for the only cases known before,
making Conjecture~\ref{conj:fk} solved for $q=0$ and $q=2$,
and Conjecture~\ref{conj:on} for $n=1$.
The method also contributes to our
understanding of universality phenomenon.

Much of the method works for general values of $n$ and $q$.
The most interesting values of the parameters
(where it does not yet work all the way)
 are $n=0$, related to 
the self-avoiding random walk, and $q=1$, 
equivalent to the critical bond percolation on the square lattice
(in the latter case some progress was achieved 
by Vincent Beffara by a different method).
Hopefully the lemma 
(essentially the discrete analyticity statement -- see below)
required to transfer our proof to other models
will be worked out someday,
leading to full resolution
of these Conjectures.

\section{Schramm-Loewner Evolution}

\subsection{Loewner Evolution}
Loewner Evolution  is a differential equation for a Riemann uniformization map
for a domain with a growing slit.
It was introduced by Charles Loewner
in \cite{Loewner} in his work  on Bieberbach's conjecture.

In the original work, Loewner considered slits growing towards interior point.
Though such  \emph{ radial} evolution 
(along with other possible setups)
is also important in the context of lattice models
and fits equally well into our framework,
we will restrict ourselves
to the   \emph{ chordal} case, when the slit is growing towards a point on the boundary.

In both cases we choose a particular Riemann map by fixing its value and derivative at the target point.
\emph{Chordal Loewner Evolution} describes uniformization
for the upper half-plane $\Cp$ with a slit growing from $0$
to $\infty$ (one deals with a general domain $\Omega$
with boundary points $a,b$ by mapping
it to $\Cp$ so that $a\mapsto0$, $b\mapsto\infty$).

Loewner only considered slits given by smooth simple curves, 
but more generally one allows any set which grows
continuously in conformal metric when viewed 
from $\infty$.
We will omit the precise definition
of \emph{allowed slits} (more extensive discussion in this context can be found in \cite{Lawler-book}), 
only noting that all simple curves 
are included.
The random curves arising from lattice models (e.g. cluster perimeters or interfaces)
are simple (or can be made simple by altering them on the local scale).
Their scaling limits are not necessarily simple, but they have no ``transversal'' self-intersections.
For such a curve to be an allowed slit it is sufficient 
if it touches itself to never venture into the created loop.
This property would follow
if e.g. a curve visits no point thrice.

Parameterizing the slit $\gamma$ in some way by time $t$, 
we denote
by $g_t(z)$ the conformal map sending
$\Cp\setminus\gamma_t$ (or rather its component at $\infty$) to $\Cp$
normalized so that at infinity
$g_t(z)=z+\alpha(t)/z+{\Cal O}(1/|z|^2)$,
the so called  \emph{ hydrodynamic normalization}.
It turns out that $\alpha(t)$ is a continuous strictly increasing
function (it is a sort of capacity-type parameter for $\gamma_t$),
so one can change the time so that
\begin{equation}
g_t(z)~=~z+\frac{2t}{z}+{\Cal O}\br{\frac{1}{|z|^2}}
\end{equation}
Denote by $w(t)$ the image of the tip $\gamma(t)$.
The family of maps $g_t$ 
(also called a  \emph{ Loewner chain}) 
is uniquely determined
by the real-valued  \emph{ ``driving term''} $w(t)$.
The general Loewner theorem can be roughly stated as follows:
\begin{lthm}
There is a bijection
between allowed slits
and continuous real valued functions $w(t)$
given by the ordinary differential equation
\begin{equation}
\partial_{t}g_t(z)~=~\frac{2}{g_t(z)-w(t)}~,
~~g_0(z)~=~z~.
\end{equation}
\end{lthm}
The original Loewner equation is different since he worked with smooth radial slits and evolved them
in another (but related) way.

\subsection{Schramm-Loewner Evolution}
While a deterministic curve $\gamma$ corresponds to a deterministic driving term $w(t)$,
a random $\gamma$ corresponds to a random $w(t)$.
One obtains \sle{\kappa} by taking $w(t)$  to be a  Brownian motion
with speed $\kappa$:
\begin{defi}
\emph{ Schramm-Loewner Evolution}, or  \sle{\kappa},
is the Loewner chain one obtains by taking $w(t)=\sqrt{\kappa}B_t$,
$\kappa\in[0,\infty)$.
Here $B_t$ denotes the standard (speed one) Brownian motion (Wiener process).
\end{defi}

The resulting slit will be almost surely
a continuous curve.
So we will also use the term \slee~ for the resulting random curve,
i.e. a probability measure on the space of curves
(to be rigorous one can think of a Borel measure on the space 
of curves with uniform norm).
Different speeds $\kappa$ produce different curves:
we grow the slit
with constant speed (measured by capacity), while the driving term 
``wiggles'' faster.
Naturally, the curves become
more ``fractal'' as $\kappa$ increases: 
for $\kappa\le4$ the curve is almost surely  simple,
for $4<\kappa<8$ it almost surely  touches itself,
and for $\kappa\ge8$ it is almost surely  space-filling
(i.e. visits every point in $\Cp$)
-- 
see \cite{Lawler-book,Rohde-Schramm}
for these and other properties.
Moreover, Vincent Beffara \cite{Beffara}
has proved that the Hausdorff dimension of the \sle{\kappa} curve is
almost surely $\min\br{1+\kappa/8,2}$.

\subsection{Conformal Markov property}
Suppose we want to describe the scaling limits of cluster perimeters, 
or interfaces for lattice models
assuming their existence and conformal invariance.
We follow Oded Schramm \cite{Schramm-lerw}
to show that Brownian motion as the driving force
arises naturally.
Consider  a simply connected domain $\Omega$ with
two boundary points, $a$ and $b$.
Superimpose a lattice with mesh $\mesh$ and
consider some lattice model,
say critical percolation with the Dobrushin boundary conditions,
leading to an interface running from $a$ to $b$,
which is illustrated by Fig.~\ref{fig:perc}
for a rectangle with two opposite corners as $a$ and $b$.
So we end up with a random simple curve (a broken line)
connecting $a$ to $b$ inside $\Omega$.
The law of the curve depends of course on the lattice superimposed.
If we believe the physicists' predictions,
as mesh tends to zero, this measure on broken lines converges 
(in an appropriate weak-$*$ topology) 
to some measure $\mu=\mu({\Omega,a,b})$
on continuous curves from $a$ to $b$ inside $\Omega$.

In this setup the conformal invariance prediction can be formulated as follows:
\begin{propa}
For a conformal map $\phi$ of the domain $\Omega$ one has
$$\phi\left(\mu(\Omega,a,b)\right)~=~\mu
\left(\phi(\Omega),\phi(a),\phi(b)\right)~.$$
\end{propa}
\begin{figure}[ht!]\centering
\unitlength=0.45mm
\def\vtx{\circle*{4}}
\begin{picture}(210,110)(0,-10)
\thicklines
\qbezier(0,20)(0,30)(10,40)
\qbezier(10,40)(20,50)(40,50)
\qbezier(40,50)(50,50)(50,20)
\qbezier(50,20)(50,0)(20,0)
\qbezier(20,0)(0,0)(0,20)
\qbezier[10](20,0)(20,20)(35,30)
\qbezier[5](35,30)(42.5,35)(40,50)
\put(20,0)\vtx
\put(40,50)\vtx
\put(11,28){$\Omega$}
\put(13,-8){$a$}
\put(33,53){$b$}
\put(6,70){${\mu(\Omega,a,b)}$}
\put(80,0)
{\qbezier(0,30)(0,30)(30,50)
\qbezier(0,30)(0,30)(20,0)
\qbezier(20,0)(20,0)(50,20)
\qbezier(30,50)(30,50)(50,20)
\qbezier[10](20,0)(20,20)(30,30)
\qbezier[5](30,30)(35,35)(30,50)
\put(20,0)\vtx
\put(30,50)\vtx
\put(25,-6){$\phi(a)$}
\put(14,53){$\phi(b)$}
\put(8,28){$\phi(\Omega)$}
\put(10,70){$\phi\br{\mu(\Omega,a,b)}$}
}
\put(170,0)
{\qbezier(0,30)(0,30)(30,50)
\qbezier(0,30)(0,30)(20,0)
\qbezier(20,0)(20,0)(50,20)
\qbezier(30,50)(30,50)(50,20)
\qbezier[10](20,0)(25,15)(30,30)
\qbezier[5](30,30)(35,25)(40,20)
\qbezier[8](40,20)(35,35)(30,50)
\put(20,0)\vtx
\put(30,50)\vtx
\put(25,-6){$\phi(a)$}
\put(14,53){$\phi(b)$}
\put(8,28){$\phi(\Omega)$}
\put(3,70){$\mu\br{\phi(\Omega),\phi(a),\phi(b)}$}
}
\put(60,70){$\longmapsto$}
\put(64,75){$\phi$}
\put(60,25){$\longrightarrow$}
\put(63,30){$\phi$}
\put(135,80){$\swarrow$}
\put(132,90){identical laws}
\put(165,80){$\searrow$}
\end{picture}
\caption{\textbf{(A) Conformal Invariance:}
conformal image of the law of the curve $\gamma$ (dotted)
in $\Omega$ coincides with the law of the curve $\gamma$ in the image domain $\phi(\Omega)$.
}
\label{fig:propa}
\end{figure}
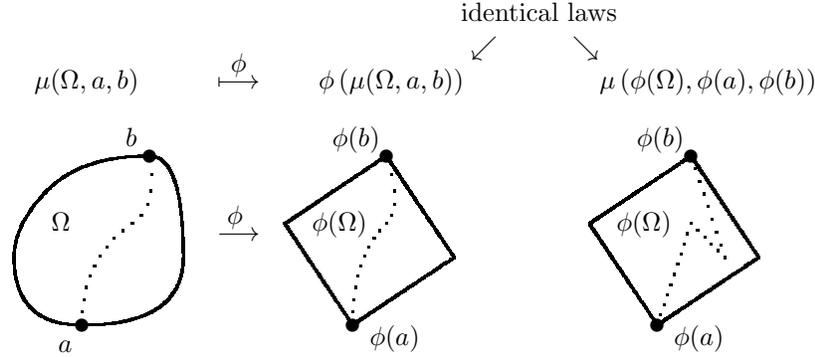
\noindent
Here a bijective map $\phi:\Omega\to\phi(\Omega)$ induces a map acting on the curves
in $\Omega$, which in turn induces a map on the probability measures on the space
of such curves, which we denote by the same letter.
By a conformal map we understand a bijection which locally preserves angles.

Moreover, if we start drawing the interface from the point $a$, we will be walking
around the grey cluster  following the right-hand rule -- see Figure~1.
If we stop at some point $a'$ after drawing the part $\gamma'$ of the interface,
we cannot distinguish the boundary of $\Omega$ from the part of the interface we have drawn:
they both are colored grey on the (counterclockwise) arc $a'b$ and white on the arc $ba'$ of
the domain $\Omega\setminus\gamma'$.
So we can say that the conditional law of the interface (conditioned on it starting as $\gamma'$)
is the same as the law in a new domain with a slit.
We expect the limit law $\mu$ to have the same property:
\begin{propb}
The law conditioned on the interface already drawn is
the same as the law in the slit domain:
$$\mu\left(\Omega,a,b\right)|\gamma'~=~\mu\left(\Omega\setminus\gamma',a',b\right)~.$$
\end{propb}
\begin{figure}[ht!]\centering
\unitlength=0.45mm
\def\vtx{\circle*{4}}
\begin{picture}(210,110)(0,-10)
\thicklines
\qbezier(0,20)(0,30)(10,40)
\qbezier(10,40)(20,50)(40,50)
\qbezier(40,50)(50,50)(50,20)
\qbezier(50,20)(50,0)(20,0)
\qbezier(20,0)(0,0)(0,20)
\qbezier[10](20,0)(20,10)(30,20)
\qbezier[12](30,20)(40,30)(40,50)
\put(11,28){$\Omega$}
\put(20,0)\vtx
\put(40,50)\vtx
\put(13,-8){$a$}
\put(33,53){$b$}
\put(6,70){${\mu(\Omega,a,b)}$}
\put(80,0)
{\qbezier(0,20)(0,30)(10,40)
\qbezier(10,40)(20,50)(40,50)
\qbezier(40,50)(50,50)(50,20)
\qbezier(50,20)(50,0)(20,0)
\qbezier(20,0)(0,0)(0,20)
\qbezier(20,0)(20,10)(30,20)
\qbezier[12](30,20)(40,30)(40,50)
\put(30,20)\vtx\put(33,13){$a'$}
\put(16,12){$\gamma'$}
\put(11,28){$\Omega$}
\put(20,0)\vtx
\put(40,50)\vtx
\put(13,-8){$a$}
\put(33,53){$b$}\put(6,70){${\mu(\Omega,a,b)}\,|\,\gamma'$}
}
\put(170,0)
{\qbezier(0,20)(0,30)(10,40)
\qbezier(10,40)(20,50)(40,50)
\qbezier(40,50)(50,50)(50,20)
\qbezier(50,20)(50,0)(21,0)
\qbezier(19,0)(0,0)(0,20)
\qbezier(19,0)(19,10.4)(29.3,20.7)
\qbezier(21,0)(21,9.6)(30.7,19.3)
\qbezier[12](30,20)(40,30)(40,50)
\put(30,20)\vtx\put(33,13){$a'$}
\put(40,50)\vtx\put(33,53){$b$}
\put(11,28){$\Omega\setminus\gamma'$}
\put(3,70){$\mu\left(\Omega\setminus\gamma',a',b\right)$}
}
\put(60,70){$\longmapsto$}
\put(45,78){conditioning}
\put(135,80){$\swarrow$}
\put(132,90){identical laws}
\put(165,80){$\searrow$}
\end{picture}
\caption{\textbf{(B) Markov property:}
The law conditioned on the curve already drawn is
the same as the law in the slit domain.
In other words when drawing the curve we do not distinguish its past from the boundary.
}
\label{fig:propb}
\end{figure}
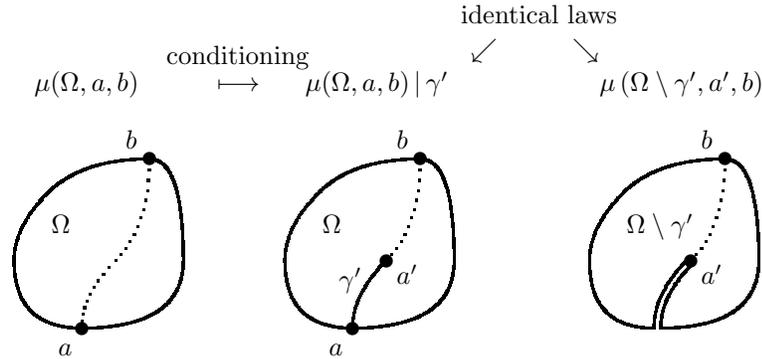

If one wants to utilize these properties to characterize $\mu$,
by (A) it is sufficient to study some reference domain (to which all others can be conformally mapped),
say the upper half-plane $\Cp$ with a curve running from $0$ to $\infty$.
Given (A), the second property (B) is easily seen to be equivalent to the following:
\begin{propc}
The law conditioned on the interface already drawn
is a conformal image of the original law.
Namely, for any conformal map $G=G_{\gamma'}$ from $\Cp\setminus\gamma'$ to $\Cp$
preserving $\infty$ and sending the tip of $\gamma'$ to $0$, we have
$$\mu\br{\Cp,0,\infty}|\gamma'~=~G^{-1}\br{\mu\br{\Cp,0,\infty}}~.$$
\end{propc}

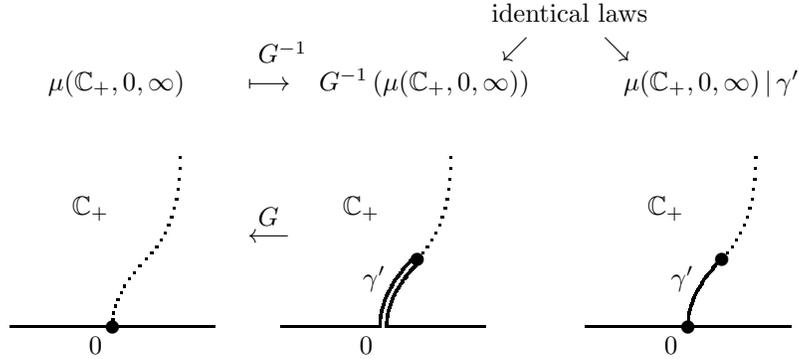
\begin{figure}[ht!]\centering
\unitlength=0.45mm
\def\vtx{\circle*{4}}
\begin{picture}(210,110)(0,-10)
\thicklines
\qbezier(-10,0)(5,0)(20,0)
\qbezier(20,0)(35,0)(50,0)
\qbezier[10](20,0)(20,10)(30,20)
\qbezier[12](30,20)(40,30)(40,50)
\put(8,33){$\Cp$}
\put(20,0)\vtx
\put(13,-8){$0$}
\put(1,70){${\mu(\Cp,0,\infty)}$}
\put(80,0)
{\qbezier(-10,0)(4.5,0)(19,0)
\qbezier(21,0)(35.5,0)(50,0)
\qbezier(19,0)(19,10.4)(29.3,20.7)
\qbezier(21,0)(21,9.6)(30.7,19.3)
\qbezier[12](30,20)(40,30)(40,50)
\put(8,33){$\Cp$}
\put(14,12){$\gamma'$}
\put(30,20)\vtx
\put(13,-8){$0$}
\put(1,70){$G^{-1}\br{\mu(\Cp,0,\infty)}$}
}
\put(170,0)
{\qbezier(-10,0)(5,0)(20,0)
\qbezier(20,0)(35,0)(50,0)
\qbezier(20,0)(20,10)(30,20)
\qbezier[12](30,20)(40,30)(40,50)
\put(8,33){$\Cp$}
\put(15,12){$\gamma'$}
\put(20,0)\vtx\put(30,20)\vtx
\put(13,-8){$0$}
\put(1,70){${\mu(\Cp,0,\infty)\,|\,\gamma'}$}
}
\put(60,70){$\longmapsto$}
\put(63,78){$G^{-1}$}
\put(60,25){$\longleftarrow$}
\put(63,30){$G$}
\put(135,80){$\swarrow$}
\put(132,90){identical laws}
\put(165,80){$\searrow$}
\end{picture}
\caption{\textbf{(B') Conformal Markov property:} 
The law conditioned on the curve already drawn
is a conformal image of the original law.
In other words the curve has ``Markov property in conformal coordinates''.
}
\label{fig:propc}
\end{figure}
\begin{rem}
Note that property (B') is formulated for the law $\mu$ for one domain only,
say $\Cp$ as above.
If we extend $\mu$ to other domains by conformal maps, it turns out that
(A) and (B) are equivalent to (B') together with scale invariance
(under maps $z\mapsto kz$, $k>0$).
\end{rem}

To use the property (B'), we describe the random curve  by the Loewner Evolution
with a certain random driving force $w(t)$
(we assume that the curve is almost surely  an allowed slit).
If we fix the time $t$, the property (B') with the slit $\gamma[0,t]$ and the map
$G_t(z)=g_t(z)-w(t)$ can be rewritten 
for random conformal map $G_{t+\delta}$ conditioned on $G_t$
(which is the same as conditioning on $\gamma[0,t]$) as
$$G_{t+\delta}|G_{t}~=~G_t\br{G_\delta}~.$$
Expanding $G$'s near infinity we obtain
$$z-w({t+\delta})+\dots|G_{t}\,=\,\br{z-w({t})+\dots}\circ\br{z-w({\delta})+\dots}\,=\,z-(w({t})+w({\delta}))+\dots\,,$$
concluding that
$$w(t+\delta)-w(t)|G_t~=~w(\delta)~.$$
This means that $w(t)$ is a continuous
(by Loewner's theorem) 
stochastic process
with independent stationary  increments.
Thus for \emph{a random curve satisfying (B')}
the driving force $w(t)$ has to be a Brownian motion
with a certain speed $\kappa\in[0,\infty)$
and drift $\alpha\in\RR$:
$$w(t)=\sqrt{\kappa} B_t+\alpha t~.$$
Applying (A) with anti-conformal reflection $\phi(u+iv)=-u+iv$ 
or with stretching $\phi(z)=2z$
shows that $\alpha$ vanishes.
So one logically arrives at the definition of \slee~and the following
\begin{sch}
A random curve satisfies (A) and (B)
if and only if it is given by \sle{\kappa} for some
$\kappa\in[0,\infty)$.
\end{sch}
The discussion above is essentially contained
in Oded Schramm's paper \cite{Schramm-lerw} for the
radial version, when slit is growing towards
a point inside and the Loewner differential
equation takes a slightly different form.
To make this principle a rigorous
statement, one has to require the curve to be almost surely an allowed slit.

\section{\slee~ as a scaling limit}

\subsection{Strategy}
In order to use the above principle 
one still has to show the  \emph{ existence} and
\emph{ conformal invariance} of the scaling limit,
and then calculate some observable
to pin down the value of $\kappa$.
For percolation one can employ its locality or
Cardy's formula for crossing probabilities
to show that $\kappa=6$.
Based on this observation
Oded Schramm concluded in \cite{Schramm-lerw}
that  \emph{if} percolation interface has a
conformally invariant scaling limit, it must be \sle6.

But it is probably difficult  to show that some interface
has a conformally invariant scaling limit
without actually identifying the latter.

To identify a random curve in principle
one needs ``infinitely many observables,''
e.g. knowing for any  finite number of
points the probability of passing above them.
This seems to be a difficult task,
which is doable for percolation since
the locality allows us
to create
many observables  from just one
(crossing probability), 
see \cite{Smirnov}.

Fortunately it turns out that even in the general case 
if an observable
has  a limit satisfying analogues of (A) and (B),
one can 
deduce convergence to \sle{\kappa}
(with $\kappa$ determined by the values of the observable).

This was demonstrated by Greg Lawler, Oded Schramm and Wendelin Werner
in \cite{Lawler-Schramm-Werner-ust}
in establishing the convergence
of two related models:
of loop erased random walk to \sle2 and
of uniform spanning tree to \sle8.

They described the discrete curve by a Loewner Evolution with
unknown random driving force.
Stopping the evolution at times $t$ and $s$
and comparing the values of the observable,
one deduces (approximate) formulae for the
conditional expectation and variance
of the increments of the driving force.
Skorokhod embedding theorem is then used
to show that driving force converges to the Brownian motion.
Finally one has to prove a (stronger) convergence
of the measures on curves.
The trick is that knowing just one observable (but \emph{for all domains})
after conditioning translates to a continuum of information
about the driving force.

We describe a different approach with the same general idea, 
which is perhaps more transparent,
separating ``exact calculations'' from ``a priori estimates''.
The idea is to first get a priori estimates, 
which imply that collection of laws is precompact
in a suitable space of allowed slits.
Then to establish the convergence it is enough to show
that limit of any converging subsequence
coincides with \slee.
To do that we describe the subsequential limit by Loewner Evolution
(with unknown random driving force $w(t)$)
and extract from the observable enough information
to evaluate expectation and quadratic variation
of increments of $w(t)$.
L\'evy's characterization implies that $w(t)$ is the Brownian motion
with a particular speed $\kappa$
and so our curves converge to \sle{\kappa}.

As an example we  discuss below an 
alternative proof of convergence to \sle6
in the case of percolation,
which uses crossing probability as an observable.
For  a domain $\Omega$ with boundary points
$a,b$ superimpose triangular lattice with mesh $\mesh$
and Dobrushin boundary conditions.
We obtain an interface $\gamma_\mesh$ 
(between open vertices on one side and closed on another)
running from $a$ to $b$,
see Fig.~\ref{fig:perc},
i.e. a measure $\mu_\mesh$ on random curves running from $a$ to $b$.

\subsection{Compactness}
First we note
that the collection $\brs{\mu_\mesh}$ is precompact (in weak-$*$
topology) in the space of continuous curves 
that are Loewner allowed slits.

The necessary framework for precompactness in the space of continuous curves
was suggested by Michael Aizenman and Almut Burchard \cite{Aizenman-Burchard}.
It turns out that appropriate bounds for probability
of an annulus being traversed $k$ times
imply tightness:
a curve has a H\"older parameterization with stochastically bounded norm.
Hence $\brs{\mu_\mesh}$ is precompact by Prokhorov's theorem:
a (uniformly controlled) part of $\mu_\mesh$
is supported on a compact set (of curves with norm bounded by $M$),
and so such parts are weakly precompact by Banach-Alaoglu theorem,
whereas the mass of the remainder tends uniformly to zero as $M\to\infty$.

The curves on the lattice are simple,
so they cannot have transversal self-intersections even after passing to the limit.
So to check that for any weak limit of $\mu_\mesh$'s
almost every curve is an allowed slit,
one has to check that as we grow it the tip is always visible 
and moves continuously when viewed
from infinity.
Essentially, one has to rule out two scenarios:
that the curve passes for a while inside already visited set,
and that the curve closes a loop, and then travels inside before exiting.
Both are reduced to probabilities of annuli traversing.

In the case of percolation one uses the Russo-Seymour-Welsh theory \cite{rsw1,rsw2}
together with Michael Aizenman's observation \cite{Aizenman-statphys} 
(that in the limit interface can visit no point thrice -- ``no $6$ arms'')
to obtain the required estimates.

Since the collection of interface laws $\brs{\mu_\mesh}$ is precompact (in weak-$*$
topology) in the space of continuous curves 
that are Loewner allowed slits,
to show that as mesh goes to zero
the interface law converge to the law of \sle6,
it is sufficient to show
that the limit of any converging subsequence is in fact \sle6.

Take some subsequence converging
to a random curve in the domain $\Omega$
from $a$ to $b$.
We map conformally to a half-plane $\Cp$, obtaining
a curve $\gamma$ from $0$ to $\infty$ with law $\mu$.
We must show that $\mu$ is given by \sle6.

By a priori estimates  $\gamma$ is almost surely  an allowed slit.
So we can describe $\gamma$ by a Loewner Evolution with 
a (random) driving force $w(t)$.
It remains to show that $w(t)=\sqrt{6}B_t$.
Note that at this point we only know that $w(t)$ is an almost surely  continuous random function
-- we do not even have a Markov property.

\subsection{Martingale observable}

Given a topological rectangle 
(a simply connected domain $\Omega$
with boundary points $a$, $b$, $c$, $d$) 
one can superimpose 
a lattice with mesh $\mesh$ onto $\Omega$
and study the probability
$\Pi_\mesh\br{\Omega,[a,b],[c,d]}$
that there is an open cluster
joining the arc $[a,b]$ to the arc $[c,d]$ on the boundary of $\Omega$.
It is conjectured that there is a limit $\Pi:=\lim_{\mesh\to0}\Pi_\mesh$,
which is conformally invariant
(depends only on the conformal
modulus of the configuration $\Omega, a, b, c, d$),
and satisfies Cardy's formula 
(predicted by John Cardy in \cite{Cardy-92} and proved in \cite{Smirnov-cras})
in half-plane:
\begin{equation}
\Pi\br{\Cp,[1-u,1],[\infty,0]}~=~\frac{\Gamma(2/3)}{\Gamma(1/3)\Gamma(4/3)}\,u^{1/3}\,{}_2F_1\br{\frac13,\frac23;\frac43;u}~=:~F(u)~.
\label{eq:cardy}
\end{equation}
Above ${}_2F_{1}$ is the hypergeometric function,
so one can alternatively write
$$F(u)~=~\left.\int_0^u\br{v(1-v)}^{-2/3}dv\,\right/\,\int_0^1\br{v(1-v)}^{-2/3}dv~.$$
Particular nature of the function is not important,
we rather use the fact that there is an explicit formula for half-plane
with four marked boundary points
and hence by conformal invariance for an arbitrary topological rectangle.
The value $\kappa=6$ will arise later from some expression involving
derivatives of $F$.

Assume that for some percolation model we are able to prove the above
conjecture
(for critical site percolation on the triangular lattice  it was proved in \cite{Smirnov-cras,Smirnov}).

Add two points on the boundary, making $\Omega$ a topological rectangle $axby$
and consider the crossing probability $\Pi_\mesh\br{\Omega,[a,x],[b,y]}$ (from the arc $ax$ to the arc $by$
on a lattice with mesh $\mesh$).

Parameterize the interface $\gamma_\mesh$ in some way by time,
and draw the part $\gamma_\mesh[0,t]$.
Note that it has open vertices on one side
(arc $\gamma_\mesh(t)a$) and closed on another
(arc $a\gamma_\mesh(t)$).
Then any open crossing from
the arc $by$ to the arc $ax$ inside $\Omega$ is
either disjoint from $\gamma_\mesh[0,t]$,
or hits its ``open'' arc $\gamma_\mesh(t)a$.
In either case it produces an  open crossing from
the arc $by$ to the arc $\gamma_\mesh(t)x$ inside $\Omega\setminus\gamma_\mesh[0,t]$,
and converse also holds.
Therefore one sees that for every realization of $\gamma_\mesh[0,t]$
the crossing probability conditioned on $\gamma_\mesh[0,t]$
coincides with crossing probability in the slit domain $\Omega\setminus\gamma_\mesh[0,t]$:
\begin{equation}\Pi_\mesh\br{\Omega,[a,x],[b,y]|\gamma_\mesh[0,t]}=
{\Pi_\mesh\br{\Omega\setminus\gamma_\mesh[0,t],[\gamma_\mesh(t),x],[b,y]}}~,
\label{eq:pp}\end{equation}
-- an analogue of the Markov property (B).
Alternatively  this follows from the fact that
$\Pi$  can be understood in terms of the interface as 
the probability that it touches the arc $xb$ before the arc $by$.
For example, on Figure~\ref{fig:perc} there is no
horizontal grey (open) crossing
(there is a vertical white crossing instead), and 
interface traced from the left upper corner $a$ touches
the lower side $xb$ before the right side $by$.

Stopping the curve at
times $t<s$ and using (\ref{eq:pp})
we can write by the total probability theorem
for every realization of $\gamma_\mesh[0,t]$
\begin{align}
\Pi_\mesh&\br{\Omega\setminus\gamma_\mesh[0,t],[\gamma_\mesh(t),x],[b,y]}\nonumber\\
&=\EE_{\gamma_\mesh[t,s]}\br{\Pi_\mesh\br{\Omega\setminus\gamma_\mesh[0,s],[\gamma_\mesh(s),x],[b,y]}|\gamma[0,t]}~.
\label{eq:complete}
\end{align}

The same a priori estimates as in the previous subsection
show that the identity (\ref{eq:complete}) also holds
for the (subsequential) scaling limit $\mu$
(strictly speaking there is an error
term in case the interface touches the arcs $[ax]$ or $[ya]$
before time $s$, but it decays very fast as we move $x$ and $y$ away from $a$).
We know that the scaling limit $\Pi:=\lim_{\mesh\to0}\Pi_{\mesh}$
of the crossing probabilities exists and is conformally invariant, so 
we can rewrite (\ref{eq:complete}) for the curve $\gamma$ 
with Loewner parameterization as
\begin{align}
\Pi&\br{\Cp\setminus\gamma[0,t],[\gamma(t),x],[\infty,y]}\nonumber\\
&=\EE_{\gamma[t,s]}\br{\Pi\br{\Cp\setminus\gamma[0,s],[\gamma(s),x],[\infty,y]}|\gamma[0,t]}~,
\label{eq:completehalf}
\end{align}
for almost every realization of $\gamma[0,t]$.
Moreover we can plug in exact values of the crossing probabilities,
given by the Cardy's formula.
Recall that the domain
$\Cp\setminus\gamma[0,t]$
is mapped to half-plane by the map $g_t(z)$
with $\gamma(t)\mapsto w(t)$.
Then the map $z\mapsto \frac{g_t(z)-g_t(y)}{g_t(x)-g_t(y)}$
also maps it to half-plane
with $\gamma(t)\mapsto \frac{w(t)-g_t(y)}{g_t(x)-g_t(y)}$, $y\mapsto0$, $x\mapsto1$.
Using conformal invariance and applying Cardy's formula we write
\begin{align}
\Pi\br{\Cp\setminus\gamma[0,t],[\gamma(t),x],[\infty,y]}
&=\Pi\br{\Cp,\left[-\frac{g_t(y)-w(t)}{g_t(x)-g_t(y)},1\right],[\infty,0]}\nonumber\\
&=F\br{\frac{g_t(x)-w(t)}{g_t(x)-g_t(y)}}~,\label{eq:mart}
\end{align}
for Cardy's hypergeometric function $F$.

\subsection{Conformally invariant martingale}\label{sec:cim}

Plugging (\ref{eq:mart}) into both sides of (\ref{eq:completehalf})
we arrive at
\begin{equation}
F\br{\frac{g_t(x)-w(t)}{g_t(x)-g_t(y)}}
=\EE_{\gamma[t,s]}
\br{F\br{\frac{g_s(x)-w(s)}{g_s(x)-g_s(y)}}|\gamma[0,t]}~.
\label{eq:completef}
\end{equation}

\begin{rem}\label{rem:martingale}
Denote by $x_t:=g_t(x)-w(t)$ and $y_t:=g_t(y)-w(t)$
trajectories of $x$ and $y$ under the random Loewner flow.
Then (\ref{eq:completef})
essentially means that $F\br{\frac{x_t}{x_t-y_t}}$ is a martingale.
\end{rem}
Since we want to extract the information about $w(t)$, 
we fix the ratio $x/(x-y):=1/3$ (anything not equal to $1/2$ would do)  
and let $x$ tend to infinity: $y:=-2x$, $x\to+\infty$.
Using the normalization $g_t(z)=z+2t/z+{\Cal O}(1/z^2)$ at infinity,
writing Taylor expansion for $F$, and plugging in values of derivatives of $F$
at $1/3$, we obtain the following expansion for the right-hand side
of (\ref{eq:completef}):
\begin{align}
\dots&=
F\br{\frac{x-w(t)+2t/x+{\Cal O}(1/x^2)}{(x+2t/x+{\Cal O}(1/x^2))-(-2x+2t/(-2x)+{\Cal O}(1/x^2))}}\nonumber \\[4pt]
&=F\br{\frac13-\frac{w(t)}{3}\frac1{x}+\frac{t}{3}\frac1{x^2}+{\Cal O}\br{\frac1{x^3}}}\nonumber\\[4pt]
&=F\br{\frac13}-\frac{w(t)}{3}F'\br{\frac13}\frac1{x}+
\br{\frac{t}{3}F'\br{\frac13}+\frac{w(t)^2}{3^2\cdot2}F''\br{\frac13}}\frac1{x^2}+{\Cal O}\br{\frac1{x^3}}\nonumber\\[4pt]
&=F\br{\frac13}-\frac1{x}\frac{\Gamma(2/3)}{\Gamma(1/3)\Gamma(4/3)}\frac{3^{1/3}}{2^{2/3}}\EE\,{w(t)}\nonumber\\[4pt]
&~~~-\frac1{x^2}\frac{\Gamma(2/3)}{\Gamma(1/3)\Gamma(4/3)} \frac1{3^{2/3}2^{5/3}}\Exp{w(t)^2-6t}
+{\Cal O}\br{\frac1{x^3}}\nonumber\\[4pt]
&=:A-\frac1{x}B\,\EE\,{w(t)}-\frac1{x^2}C\,\Exp{w(t)^2-6t}
+{\Cal O}\br{\frac1{x^3}}~,\nonumber
\end{align}
where we plugged in values of the derivative for hypergeometric function.
Using similar reasoning for the right-hand side of (\ref{eq:completef})
we arrive at the following identity:
\begin{align}
A&-\frac1{x}B\,\EE\,w(t)-\frac1{x^2}C\,\Exp{w(t)^2-6t}+{\Cal O}\br{\frac1{x^3}}\nonumber\\
&=A-\frac1{x}B\,\EE_{\gamma[t,s]}\br{w(s)|\gamma[0,t]}
-\frac1{x^2}C\,\EE_{\gamma[t,s]}\br{w(s)^2-6s|\gamma[0,t]}
+{\Cal O}\br{\frac1{x^3}}.\nonumber
\label{eq:completei}
\end{align}
Equating coefficients in the series above,
we conclude
that
\begin{equation}
\EE_{w[t,s]}\br{w(s)|w[0,t]}=0~,~~~
\EE_{w[t,s]}\br{w(s)^2-6s|w[0,t]}=w(t)^2-6t~.
\label{eq:levy}
\end{equation}
Thus $w(t)$ is a continuous (by Loewner's theorem)
process such that both 
$$w(t)\mathrm{~~~and~~~}w(t)^2-6t~$$ 
are 
martingales so
by L\'evy's characterization of the Brownian motion $w(t)=\sqrt{6}B_t$,
and therefore \sle6 is the scaling limit of the critical percolation interface.

The argument will work wherever Cardy's formula and a priori estimates are available,
particularly for triangular lattice.
More generally, any conformally invariant
martingale  will do, with value of $\kappa$ arising from its Taylor expansion.

\begin{rem}\label{rem:calcul}
The scheme can also be reversed to do calculations for \slee's,
if an observable is a martingale (e.g. crossing probability).
Indeed, writing the same formulae with $x/(x-y)=a$ 
we conclude that the coefficient by $1/x^2$, namely
$$C=\frac{2a(1-2a)}{1-a}tF'(a)+\frac{a}{2}\Exp{w(t)^2}F''(a)~,$$
vanishes.
Since for $w(t)=\sqrt6B(t)$ one has $\Exp{w(t)^2}=6t$,
we arrive at the differential equation
$$\frac{2(1-2a)}{3(1-a)}F'(a)+F''(a)=0.$$
With the given boundary data
it has a unique solution, which is 
Cardy's hypergeometric function.
 \end{rem}

\section{Ising model and beyond}

The martingale method as described above shows that to 
construct a conformally invariant scaling limit
for some model we need a priori estimates
and a non-trivial martingale observable with a conformally invariant scaling limit.

\subsection{A priori estimates}
A priori estimates are necessary to show that collection of interface laws
is precompact in weak-$*$ topology (on the space of
measures on continuous curves which are allowed slits).

If we follow the same route as for percolation
(via  the work \cite{Aizenman-Burchard}
of Michael Aizenman and Almut Burchard),
we only need 
to evaluate probabilities of 
traversals of an annulus in terms of its modulus.
For percolation such estimates are (almost) readily
available from the Russo-Seymour-Welsh theory.
For  uniform spanning tree and loop erased random walk one can derive
the estimates using random walk connection
and the known estimates for the latter
(a ``branch'' of a uniform spanning tree is a loop erased random walk),  
see \cite{ABNW,Schramm-lerw}.

For the Ising model the required estimates do not seem to be readily available, 
but  a vast arsenal of methods is at hand. 
Essentially all we need 
can be reduced by monotonicity arguments 
to spin correlation estimates
of Bruria Kaufman, Lars Onsager and Chen Ning Yang \cite{Kaufman-Onsager,Yang}.

For general random cluster or loop models 
such exact results are not available, but we
actually need much weaker statements,
and many of the techniques used by us for the Ising model
(like FKG inequalities) are well-known
in the general case.

So this part does not seem to be the main obstacle
to construction of scaling limits,
though it might require very hard work.
Moreover, following the proposed approach we actually get
that interfaces have a H\"older parameterization with uniformly stochastically bounded norm.
Thus rather weak kinds of convergence of interfaces would lead to convergence
in uniform norm (or rather weak-$*$ convergence
of measures on curves with uniform norm).

It also appears that the same a priori estimates can be 
employed to show
observable convergence 
in the cases concerned,
and hopefully they will be sufficient for other models.
So a more pressing question is how to construct a
martingale observable.

\subsection{Conformally covariant martingales}

Suppose that for every simply connected
domain $\Omega$
with a boundary point $a$
we have defined a random curve $\gamma$ starting from $a$.
Mark several points $b,c,\dots$ in $\Omega$ or on the boundary.
Remark~\ref{rem:martingale}
suggests the following definition:
\begin{defi}\label{def:ccm}
We say that a function  (or rather a \emph{differential})
$F(\Omega,a,b,c,\dots)$
is a \emph{conformal (covariant) martingale}
for a random curve $\gamma$
if 
\begin{align}
&F\mathrm{~is~conformally~covariant:~~}
F\left(\Omega,a,b,c,\dots\right)
=\nonumber\\
=&F\left(\phi(\Omega),\phi(a),\phi(b),\phi(c),\dots\right)
\cdot\phi'(b)^\alpha\bar\phi'(b)^{\beta}\phi'(c)^\gamma\bar\phi'(c)^{\delta}\dots,
\label{eq:cov}
\end{align}
and 
\begin{equation}\label{eq:m}
F(\Omega\setminus\gamma[0,t],\gamma(t),b,c,\dots)
\mathrm{~~~is~a~martingale}
\end{equation}
with respect to the random
curve $\gamma$ drawn from $a$ (with Loewner parameterization).
\end{defi}
Introducing covariance at $b,c,\dots$ 
we do not  ask for covariance at $a$,
since it always can be rewritten as covariance at other points.
And applying factor at $a$ would be troublesome: 
once we started drawing a curve the domain becomes non-smooth in its neighborhood,
creating problems with the definition.

If the exponents $\alpha,\beta,\dots$ vanish, we obtain an invariant quantity.
While the crossing probability for the percolation was invariant,
many quantities of interest in physics
are covariant differentials,
e.g. cluster density at $c$ would scale as a lattice step to some
power (depending on the model),
so we would arrive at a factor
$$|\phi'(c)|^\delta=\phi'(c)^{\delta/2}\bar\phi'(c)^{\delta/2}~.$$
There are other possible generalizations, e.g. 
one can add the Schwarzian derivative of $\phi$ to (\ref{eq:cov}).
The two properties in Definition~\ref{def:ccm} are analogues 
of (A) and (B), and similarly combined
they show that for the curve $\gamma$ mapped to half-plane
from any domain $\Omega$ so that $a\mapsto0$, $b\mapsto\infty$, $c\mapsto x$ 
(note that the image curve in $\Cp$ might depend on $\Omega$
-- we only know  the conformal invariance of an observable,
not of the curve itself)
we have an analogue of (B'),
which was already mentioned in Remark~\ref{rem:martingale}
for percolation.
Namely
$$F\br{\Cp,0,\infty,g_t(x),\dots}\cdot
g_t'(x)^\gamma\bar g_t'(x)^{\delta}\dots,$$
is a martingale with respect to the
random Loewner evolution
(covariance factor at $b=\infty$ is absent,
since $g_t'(\infty)=1$).

The equation (\ref{eq:completef})
can be written
for this $F$, and if we can evaluate $F$ exactly, the same machinery 
as one used by Greg Lawler, Oded Schramm and Wendelin Werner
in \cite{Lawler-Schramm-Werner-ust}
or as the one discussed above for percolation proves that our random curve is \slee.
So one arrives at a following generalization of Oded Schramm's
principle:

\begin{pri}
If a random curve $\gamma$ admits a (non-trivial) conformal
martingale $F$, then $\gamma$ is given by \slee~with 
$\kappa$ (and drift depending on modulus of the configuration)
derived from $F$.
\end{pri}
\begin{rem}
In chordal situation we consider curves
growing from $a$ towards another boundary point $b$ in a simply connected domain.
But the same conclusion would hold on general domains or Riemann surfaces with boundary
once we find a covariant martingale
(for appropriate generalizations of Loewner Evolutions 
see e.g. the book \cite{A-book}).
The only difference is that driving force of the corresponding Loewner Evolution
will be a Brownian motion with drift depending on conformal
modulus of the configuration $\Omega,a,b,c,\dots$,
leading to \slee~generalizations.
Starting from lattice models with various boundary conditions
and conditioned on various events, one can see which drifts
will be of interest for \slee~generalizations.
\end{rem}

\subsection{Discrete analyticity}

Passing to the lattice model, we
want to find a discrete object, which
in the limit becomes a conformally covariant martingale.

Martingale property 
is actually more accessible in the discrete setting.
For example, functions which are defined
as observables (like probability of the interface going through a vertex,
edge density for the model, etc.)
have the martingale property built in,
and so only conformal covariance must be established.

Alternatively, one can work with a discrete function
$F(\Omega,a,b,c)$ 
(a priori not related to lattice models)
which has a conformally covariant 
scaling limit by construction.
Then we need to connect it to a particular lattice model,
establishing a martingale property (\ref{eq:m}).
In the discrete case it is sufficient to check the latter
for a curve advanced by one step.
Assume that once we have drawn the part $\gamma'$
of the interface
from the point $a$ to point $a'$,
it turns left with probability $p=p(\Omega,\gamma',a',b)$
creating a curve $\gamma_l=\gamma\cup\brs{a_l}$
or right with probability $(1-p)$ creating
a curve $\gamma_r=\gamma\cup\brs{a_r}$.
Then it is enough to check the identity
\begin{align}\label{eq:reverse}
F(\Omega\setminus\gamma',a',b,c)=pF(\Omega\setminus\gamma_l,a_l,b,c)+(1-p)F(\Omega\setminus\gamma_r,a_r,b,c)~,\\
p=p(\Omega,\gamma',a',b)~.\nonumber
\end{align}
Actually our proof for the Ising model can be rewritten that
way, with $F$ defined as a solution of an appropriate discretization 
of the Riemann Boundary Value Problem (\ref{eq:rbvp})
-- the observable nature of $F$ never comes up.

Moreover, starting with $F$ one can define a random curve
by choosing ``turning probabilities'' $p$ so that
identity (\ref{eq:reverse}) is satisfied,
obtaining a model with conformally invariant scaling limit
by ``reverse engineering.''
For example, starting with a harmonic function of $c$ with
boundary values $1$ on the arc $ba$ and $0$ on the arc $ab$,
one obtains a unique discrete random curve, which has it as a martingale.
Note that such a function is a particular case $\alpha=1$ of the martingale (\ref{eq:phi})
below,  corresponding to $\kappa=4$
(or rather its integral).
In \cite{Schramm-Sheffield}  Oded Schramm and Scott Sheffield 
 introduced this curve with a nicer ``Harmonic Explorer'' definition,
and utilizing the mentioned observable showed that it indeed converges to \sle4.
It seems that in this way one can use the solutions to the problem (\ref{eq:rbvp})
to construct models converging to arbitrary \slee's, however 
it is not clear though whether they would similarly have ``nicer'' definitions.

Anyway, for either approach to work
we need a
\emph{discrete conformal covariant}
with a scaling limit.
We have tried discretizations of many conformally invariant objects
(extremal length, capacity, solutions to variational problems, \dots)
and the most promising in this context seem to be discrete
harmonic or analytic functions
in additional variable(s) (in $c,\dots$).
Firstly, all other invariants can be rewritten in this way.
Secondly, discretization of harmonic and analytic functions is a nice 
and very well studied (especially in the case of harmonic ones) object.
Thirdly, one can obtain very non-trivial
invariants by just checking local conditions:
harmonicity or analyticity inside plus some boundary conditions
(Dirichlet, Neumann, Riemann-Hilbert, etc.).
The most natural candidate would be a harmonic function solving some Dirichlet
problem. 

Note that such an observable is known for the Brownian motion.
A classical theorem \cite{Kakutani-44} of Shizuo Kakutani
states that in a domain $\Omega$ exit probabilities for Brownian motion
started at $z$ are harmonic functions in $z$ with
easily determinable boundary values.
Though Kakutani works directly with Brownian motion,
one can do the same for the random walk
(which is actually much easier,
since discrete Laplacian of the exit probability is trivially zero),
and then passing to a limit
deduce statements about Brownian motion,
including its conformal invariance.

\subsection{Classification of conformal martingales}

Before we start working in the discrete setting,
we might want to investigate which functions 
are conformal martingales for \slee~curves, 
and so can arise
as scaling limits of martingale observables
for lattice models. 

As discussed in Remark~\ref{rem:calcul},
one can write partial differential equations
for \slee~ conformal martingales.
For small number of points those equations
can be solved,
and in such a way one computes dimensions, scaling exponents and other quantities of interest. 
For any particular value of $\kappa$
we can see which martingales have the simplest form
and so  are probably easier to work with.
Also if they have a geometric \slee~ interpretation
(like probability of \slee~curve going to one side
of a point, etc.)
we can study similar quantities for the lattice model.

It turns out that only  for $\kappa=4$
one obtains a nice harmonic martingale with Dirichlet boundary conditions.
In that case  the probability of \sle4 passing to one side of a point $z$
is harmonic in $z$ and has boundary conditions $0$ and $1$, see
 \cite{Schramm-formula}.
Oded Schramm and Scott Sheffield \cite{Schramm-Sheffield}
constructed a model
which has this property on discrete level built in. 
Unfortunately the property was not yet
observed in any of the classical models conjecturally converging
to \sle4, though results of Kenyon \cite{Kenyon1}
show it holds for double-domino curves
in Temperley domains (i.e.  a domain with the boundary
satisfying a certain local condition).

In the case of mixed Dirichlet-Neumann conditions, 
it becomes possible to work with some other values of $\kappa$,
including  uniform spanning tree $\kappa=8$, which 
is exploited in \cite{Lawler-Schramm-Werner-ust}.
There are also covariant candidates for a few other values of $\kappa$
(notably $8/3$ which corresponds to self-avoiding random walk),
but they were not yet observed in lattice models.

Thus to study general models, one is forced to utilize more
general boundary value problems
with a Riemann(-Hilbert) Boundary Value Problem
being the natural candidate.
Besides harmonic function it involves its harmonic conjugate,
and so is better formulated in terms of analytic functions.
Moreover, discrete analyticity involves a first order Cauchy-Riemann operator,
rather than a second order Laplacian,
and  so it should be easier to deal with than harmonicity.

As discussed above, we can classify all analytic martingales.
For chordal \slee~ and
$F$ with three points $a,b,z$ as parameters
we discover two particularly nice families.
The following Proposition will be discussed in \cite{smirnov-fk2,smirnov-is}
and our subsequent work:
\begin{prop}\label{prop:mart}
Let $\Omega$ be a simply connected domain with boundary points $a$, $b$.
Let $\Phi(z)=\Phi(\Omega,a,b,z)$ be a mapping of $\Omega$ to a horizontal strip $\RR\times[0,1]$,
such that $a$ and $b$ are mapped to $\mp\infty$.
Then 
\begin{equation}\label{eq:phi}
F(\Omega,a,b,z)=\Phi'(z)^{\alpha}{\mathrm{~~with~~}}\alpha=\frac8\kappa-1~,
\end{equation}
is a martingale for \sle{\kappa}.
Let $\Psi(z)=\Psi(\Omega,a,b,z)$ be a mapping of $\Omega$ to a half-plane $\Cp$,
such that $a$ and $b$ are mapped to $\infty$ and $0$ correspondingly.
Then 
\begin{equation}\label{eq:psi}
G(\Omega,a,b,z)=\Psi'(z)^{\alpha}\Psi'(b)^{-\alpha}{\mathrm{~~with~~}}\alpha=\frac3\kappa-\frac12~,
\end{equation}
 is a martingale for \sle{\kappa}.
\end{prop}

These martingales make most sense for
$\kappa\in[4,8]$ and $\kappa\in[8/3,8]$ correspondingly,
and are related to observables of interest
in Conformal Field Theory
(which was part of our motivation to introduce them).
Note that both functions are covariant with power $\alpha$
(which is the spin in physics terminology), and solve the
Riemann boundary value problems
\begin{equation}\label{eq:rbvp}
\IM\br{F(z)\tau(z)^\alpha}=0,~~z\in\partial\Omega,
\end{equation}
where $\tau(z)$ is the tangent vector to $\partial\Omega$ at $z$.

The problem is to observe these functions in the discrete setting,
and some intuition can be obtained from their geometric meaning for \slee's.
For example, $F$ is roughly speaking
(one has to consider an intermediate scale to make sense of it)
an expectation of \slee~curve passing
through $z$ taken with some complex weight depending
on the winding.

\subsection{Height models and Coulomb gas}

The above-mentioned expectation actually
makes more sense
(and is immediately well-defined)
 in the discrete setting
and one arrives at the same object with the same complex weight
via several different approaches.

One way is to consider the Coulomb gas arguments
(cf. \cite{N2} by Bernard Nienhuis) for the loop representation.
In the random cluster case at criticality,
the weight of a loop is $\sqrt{q}$ -- recall (\ref{eq:fk}). 
We randomly and independently orient the loops,
and introduce the height function $h$
which whenever a loop is crossed changes by $\pm1$ 
(depending on loop direction -- think of a topographic map).
One could weight oriented loops by $\sqrt{q}/2$, obtaining essentially the same model.
However it makes sense to consider a complex weight instead.
When $q$ is in the $[0,4]$ range, there is
a complex unit number $\mu=\exp(k\cdot2\pi i)$ such that
\begin{equation}\label{eq:k}
k=\frac1{{2\pi}}{\arccos\br{\sqrt{q}/2}}\mathrm{~~or~~}\mu+\bar\mu=\sqrt{q}.
\end{equation}
We (independently and randomly) orient all loops,
prescribing weight $\mu$ per counterclockwise
and $\bar\mu$ per clockwise loop.

Forgetting orientation of loops
reconstructs the original model.
Unfortunately the new partition function is complex and no longer leads
to a probability measure (moreover, its variation blows up as the lattice step goes to zero),
but it can be defined locally,
making it much more accessible.

Indeed, going around a cycle, 
and turning by $\Delta_z$
at vertex $z$,
the total sum of turns $\sum_{z\in\mathrm{cycle}}\Delta_z$
is $\pm2\pi$ depending on whether the cycle
is counter or clockwise.
So the weight per cycle can be written as
$\prod_{z\in\mathrm{cycle}}\exp(ik\cdot\Delta_z)$
and the total weight of the configuration is 
$\prod_{z\in\Omega}\exp(ik\cdot\Delta_z)$,
which can be computed locally 
(without reference to the global order of cycles).
The same weight can also be written in terms of the gradient of height function.

The interface is always oriented from $a$ to $b$,
so that the height function is always equal to $0$
on the arc $ab$ and to $1$ on the arc $ba$.
From physics arguments the interface curve 
(being ``attached'' to the boundary on both sides)
should be weighted differently from loops, 
namely by $\exp(i(2k-1/2)\cdot\Delta_z)$
per turn.
When interface runs 
between two boundary points
(being oriented from $a$ to $b$),
these factors do not matter, since total turn
from $a$ to $b$ is independent of the configuration.

However, if we choose a point $z$ on an interface
and reverse the orientation of one of its halves
(so that it is oriented from $a$ to $z$
and from $z$ to $b$),
the interface inputs a non-constant complex factor.
This orientation reversal has a nice meaning:
after it the height function acquires a $+2$ monodromy at $z$:
when we go around $z$ we cross two curves 
(halves of the interface) incoming into it.

All the loops (when we forget their orientation) still contribute the same $\sqrt{q}$
~per loop, and the complex weight can be expressed 
in terms of the interface winding (total turn expressed in radians)
from $b$ to $z$, denoted by $\wi(\gamma,b\to z)$.
So one logically arrives at
the partition function $Z$ for our model with $+2$ monodromy at $z$:
\begin{equation}\label{eq:fdef}
F(\Omega,a,b,z):=Z_{+2\mathrm{~monodromy~at~} z}=\EE\chi_{z\in\gamma}\exp\br{i(4k-1) \wi(\gamma,b\to z)}.
\end{equation}
This function is clearly a martingale,
and there are strong indications 
(both from mathematics and physics points of view)
that it is discrete analytic.

This follows from the fact that
the interface can arrive at a boundary point $z$
from $b$ with a unique winding equal to the winding of
the boundary from $b$ to $z$,
so we can express it in terms of the tangent vector $\tau(z)$.
Writing this down, we discover
that the function $F$ solves a discrete version
of the Riemann Boundary Value Problem (\ref{eq:rbvp})
with $\alpha=1-4k$.
\begin{rem}
The continuum problem was solved by the function
(\ref{eq:phi}),
so if we establish discrete analyticity
it only remains to show that a solution
to a discrete Riemann Boundary Value Problem converges to its continuum counterpart.
Moreover, combining identity $\alpha=1-4k$ with (\ref{eq:phi})
and (\ref{eq:k})
we obtain the relation between
$\kappa$ and $q$ stated in Conjecture~\ref{conj:fk}.
\end{rem}

This convergence problem seems to be difficult (and open in the general case).
The way we solve it in the Ising case is sketched below.

There are other indications that this function is nice to work with.
Indeed, the easiest form of discrete analyticity
involves local partial difference relations,
and to prove those we should count configurations
included into our expectation.
To obtain relations, we need some bijections in the configuration space,
and the easiest ones are given by local rearrangements
(we worked with global rearrangements for percolation \cite{Smirnov-cras},
but such work must be more difficult for non-local models).

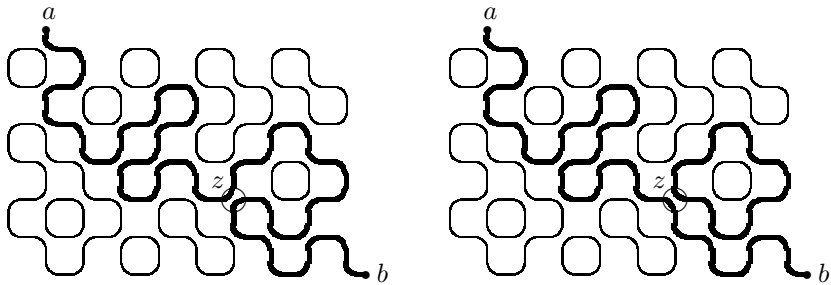
\begin{figure}\centering
{\def\bs{\circle*{3}}\def\ws{\circle{3}}
\def\ne{\qbezier(-0.,3)(0.5,0.5)(3,0)}
\def\nw{\qbezier(-0.,3)(-0.5,0.5)(-3,0)}
\def\se{\qbezier(-0.,-3)(0.5,-0.5)(3,0)}
\def\sw{\qbezier(-0.,-3)(-0.5,-0.5)(-3,0)}
\def\bne{\qbezier(-0.5,3)(0.,0.5)(3,-0)}
\def\bnw{\qbezier(-0.5,3)(-1,0.5)(-3,0)}
\def\bse{\qbezier(-0.5,-3)(0.,-0.5)(3,0)}
\def\bsw{\qbezier(-0.5,-3)(-1,-0.5)(-3,0)}
\def\uu{{\thicklines\line(1,1){10}}}
\def\dd{{\thicklines\line(1,-1){10}}}
\def\uw{\line(1,1){10}}
\def\dw{\line(1,-1){10}}
\unitlength=0.5mm
\centerline{
\begin{picture}(115,85)(-10,-10)
{\linethickness{1.5pt}
\put(10,65){\line(0,-1){2}}\put(10,47){\line(0,-1){4}}\put(20,37){\line(0,-1){4}}\put(20,57){\line(0,-1){4}}
\put(30,37){\line(0,-1){4}}\put(30,27){\line(0,-1){4}}\put(40,47){\line(0,-1){4}}\put(40,37){\line(0,-1){4}}
\put(40,27){\line(0,-1){4}}\put(50,47){\line(0,-1){4}}\put(50,27){\line(0,-1){4}}\put(60,27){\line(0,-1){4}}
\put(60,17){\line(0,-1){4}}\put(70,37){\line(0,-1){4}}\put(70,17){\line(0,-1){4}}\put(70,7){\line(0,-1){4}}
\put(80,37){\line(0,-1){4}}\put(80,17){\line(0,-1){4}}\put(80,7){\line(0,-1){4}}\put(90,27){\line(0,-1){4}}
\put(90,07){\line(0,-1){4}}\put(73,0){\line(1,0){4}}\put(63,10){\line(1,0){4}}\put(83,10){\line(1,0){4}}
\put(33,20){\line(1,0){4}}\put(53,20){\line(1,0){4}}\put(63,20){\line(1,0){4}}\put(83,20){\line(1,0){4}}
\put(93,00){\line(1,0){2}}\put(23,30){\line(1,0){4}}\put(33,30){\line(1,0){4}}\put(63,30){\line(1,0){4}}
\put(83,30){\line(1,0){4}}\put(13,40){\line(1,0){4}}\put(33,40){\line(1,0){4}}\put(43,40){\line(1,0){4}}
\put(73,40){\line(1,0){4}}\put(13,50){\line(1,0){4}}\put(43,50){\line(1,0){4}}\put(13,60){\line(1,0){4}}
\put(43,30){\line(1,0){4}}\put(73,10){\line(1,0){4}}
\put(10,60)\bne\put(10,40)\bne\put(20,30)\bne\put(30,20)\bne\put(50,20)\bne\put(60,10)\bne
\put(70,00)\bne\put(70,10)\bne\put(80,30)\bne
\put(20,50)\bnw\put(30,30)\bnw\put(40,20)\bnw\put(40,40)\bnw\put(50,40)\bnw\put(60,20)\bnw
\put(70,30)\bnw\put(80,00)\bnw\put(90,10)\bsw\put(90,20)\bnw\put(80,10)\bnw\put(40,30)\bnw
\put(20,60)\bsw\put(20,40)\bsw\put(50,30)\bsw\put(50,50)\bsw\put(70,10)\bsw\put(70,20)\bsw
\put(80,40)\bsw\put(90,30)\bsw
\put(10,50)\bse\put(30,30)\bse\put(30,40)\bse\put(40,30)\bse\put(40,40)\bse\put(40,50)\bse
\put(60,20)\bse\put(60,30)\bse\put(70,40)\bse\put(80,10)\bse\put(80,20)\bse\put(90,0)\bne
}
\put(10,00)\ne\put(30,00)\ne\put(50,00)\ne\put(00,10)\ne\put(40,10)\ne\put(50,10)\ne
\put(20,20)\ne\put(70,20)\ne\put(00,30)\ne\put(10,30)\ne\put(50,30)\ne\put(20,40)\ne
\put(80,40)\ne\put(00,50)\ne\put(30,50)\ne\put(50,50)\ne\put(60,50)\ne\put(70,50)\ne
\put(80,50)\ne\put(10,10)\ne
\put(20,00)\nw\put(40,00)\nw\put(60,00)\nw\put(20,10)\nw\put(30,10)\nw\put(10,20)\nw
\put(80,20)\nw\put(60,30)\nw\put(30,40)\nw\put(70,40)\nw\put(90,40)\nw\put(10,50)\nw
\put(40,50)\nw\put(60,40)\nw
\put(10,10)\sw\put(40,10)\sw\put(50,10)\sw\put(60,10)\sw\put(20,20)\sw\put(30,20)\sw
\put(50,20)\sw\put(10,30)\sw\put(20,30)\sw\put(80,30)\sw\put(10,40)\sw\put(60,50)\sw
\put(70,50)\sw\put(80,50)\sw\put(90,50)\sw\put(10,60)\sw\put(40,60)\sw\put(60,60)\sw
\put(80,60)\sw\put(30,50)\sw
\put(20,10)\se\put(30,10)\se\put(00,20)\se\put(10,20)\se\put(40,20)\se\put(70,30)\se
\put(00,40)\se\put(50,40)\se\put(20,50)\se\put(00,60)\se\put(30,60)\se\put(60,40)\se
\put(50,60)\se\put(70,60)\se
\put(00,17){\line(0,-1){4}}\put(00,37){\line(0,-1){4}}\put(00,57){\line(0,-1){4}}\put(10,07){\line(0,-1){4}}
\put(10,17){\line(0,-1){4}}\put(10,27){\line(0,-1){4}}\put(10,37){\line(0,-1){4}}\put(10,47){\line(0,-1){4}}
\put(20,07){\line(0,-1){4}}\put(20,17){\line(0,-1){4}}\put(20,27){\line(0,-1){4}}\put(20,47){\line(0,-1){4}}
\put(30,07){\line(0,-1){4}}\put(30,17){\line(0,-1){4}}\put(30,47){\line(0,-1){4}}\put(30,57){\line(0,-1){4}}
\put(40,07){\line(0,-1){4}}\put(40,17){\line(0,-1){4}}\put(40,57){\line(0,-1){4}}\put(50,07){\line(0,-1){4}}
\put(50,17){\line(0,-1){4}}\put(50,37){\line(0,-1){4}}\put(50,57){\line(0,-1){4}}\put(60,07){\line(0,-1){4}}
\put(60,37){\line(0,-1){4}}\put(60,57){\line(0,-1){4}}\put(60,47){\line(0,-1){4}}\put(70,27){\line(0,-1){4}}
\put(70,47){\line(0,-1){4}}\put(70,57){\line(0,-1){4}}\put(80,27){\line(0,-1){4}}\put(80,47){\line(0,-1){4}}
\put(80,57){\line(0,-1){4}}\put(90,47){\line(0,-1){4}}\put(10,57){\line(0,-1){4}}
\put(13,00){\line(1,0){4}}\put(33,00){\line(1,0){4}}\put(53,00){\line(1,0){4}}\put(03,10){\line(1,0){4}}
\put(13,10){\line(1,0){4}}\put(23,10){\line(1,0){4}}\put(33,10){\line(1,0){4}}\put(43,10){\line(1,0){4}}
\put(53,10){\line(1,0){4}}\put(03,20){\line(1,0){4}}\put(13,20){\line(1,0){4}}\put(23,20){\line(1,0){4}}
\put(43,20){\line(1,0){4}}\put(03,30){\line(1,0){4}}\put(13,30){\line(1,0){4}}\put(53,30){\line(1,0){4}}
\put(53,40){\line(1,0){4}}\put(63,40){\line(1,0){4}}\put(83,40){\line(1,0){4}}\put(03,50){\line(1,0){4}}
\put(03,40){\line(1,0){4}}\put(23,40){\line(1,0){4}}\put(73,30){\line(1,0){4}}\put(73,20){\line(1,0){4}}
\put(23,50){\line(1,0){4}}\put(33,50){\line(1,0){4}}\put(53,50){\line(1,0){4}}\put(63,50){\line(1,0){4}}
\put(73,50){\line(1,0){4}}\put(83,50){\line(1,0){4}}\put(03,60){\line(1,0){4}}\put(33,60){\line(1,0){4}}
\put(53,60){\line(1,0){4}}\put(73,60){\line(1,0){4}}
\put(09,68){$a$}\put(98,-2){$b$}\put(54,23){$z$}
\put(10,65){\circle*{2}}
\put(60,20){\circle{6}}
\put(95,00){\circle*{2}}
\end{picture}
\begin{picture}(115,85)(-10,-10)
{\linethickness{1.5pt}
\put(10,65){\line(0,-1){2}}\put(10,47){\line(0,-1){4}}\put(20,37){\line(0,-1){4}}\put(20,57){\line(0,-1){4}}
\put(30,37){\line(0,-1){4}}\put(30,27){\line(0,-1){4}}\put(40,47){\line(0,-1){4}}\put(40,37){\line(0,-1){4}}
\put(40,27){\line(0,-1){4}}\put(50,47){\line(0,-1){4}}\put(50,27){\line(0,-1){4}}\put(60,27){\line(0,-1){4}}
\put(60,17){\line(0,-1){4}}\put(70,37){\line(0,-1){4}}\put(70,17){\line(0,-1){4}}\put(70,7){\line(0,-1){4}}
\put(80,37){\line(0,-1){4}}\put(80,17){\line(0,-1){4}}\put(80,7){\line(0,-1){4}}\put(90,27){\line(0,-1){4}}
\put(90,07){\line(0,-1){4}}\put(73,0){\line(1,0){4}}\put(63,10){\line(1,0){4}}\put(83,10){\line(1,0){4}}
\put(33,20){\line(1,0){4}}\put(53,20){\line(1,0){4}}\put(63,20){\line(1,0){4}}\put(83,20){\line(1,0){4}}
\put(93,00){\line(1,0){2}}\put(23,30){\line(1,0){4}}\put(33,30){\line(1,0){4}}\put(63,30){\line(1,0){4}}
\put(83,30){\line(1,0){4}}\put(13,40){\line(1,0){4}}\put(33,40){\line(1,0){4}}\put(43,40){\line(1,0){4}}
\put(73,40){\line(1,0){4}}\put(13,50){\line(1,0){4}}\put(43,50){\line(1,0){4}}\put(13,60){\line(1,0){4}}
\put(43,30){\line(1,0){4}}\put(73,10){\line(1,0){4}}
\put(10,60)\bne\put(10,40)\bne\put(20,30)\bne\put(30,20)\bne\put(50,20)\bne\put(60,10)\bne
\put(70,00)\bne\put(70,10)\bne\put(80,30)\bne
\put(20,50)\bnw\put(30,30)\bnw\put(40,20)\bnw\put(40,40)\bnw\put(50,40)\bnw\put(60,20)\bne
\put(70,30)\bnw\put(80,00)\bnw\put(90,10)\bsw\put(90,20)\bnw\put(80,10)\bnw\put(40,30)\bnw
\put(20,60)\bsw\put(20,40)\bsw\put(50,30)\bsw\put(50,50)\bsw\put(70,10)\bsw\put(70,20)\bsw
\put(80,40)\bsw\put(90,30)\bsw
\put(10,50)\bse\put(30,30)\bse\put(30,40)\bse\put(40,30)\bse\put(40,40)\bse\put(40,50)\bse
\put(60,20)\bsw
\put(60,30)\bse\put(70,40)\bse\put(80,10)\bse\put(80,20)\bse\put(90,0)\bne
}
\put(10,00)\ne\put(30,00)\ne\put(50,00)\ne\put(00,10)\ne\put(40,10)\ne\put(50,10)\ne
\put(20,20)\ne\put(70,20)\ne\put(00,30)\ne\put(10,30)\ne\put(50,30)\ne\put(20,40)\ne
\put(80,40)\ne\put(00,50)\ne\put(30,50)\ne\put(50,50)\ne\put(60,50)\ne\put(70,50)\ne
\put(80,50)\ne\put(10,10)\ne
\put(20,00)\nw\put(40,00)\nw\put(60,00)\nw\put(20,10)\nw\put(30,10)\nw\put(10,20)\nw
\put(80,20)\nw\put(60,30)\nw\put(30,40)\nw\put(70,40)\nw\put(90,40)\nw\put(10,50)\nw
\put(40,50)\nw\put(60,40)\nw
\put(10,10)\sw\put(40,10)\sw\put(50,10)\sw\put(60,10)\sw\put(20,20)\sw\put(30,20)\sw
\put(50,20)\sw\put(10,30)\sw\put(20,30)\sw\put(80,30)\sw\put(10,40)\sw\put(60,50)\sw
\put(70,50)\sw\put(80,50)\sw\put(90,50)\sw\put(10,60)\sw\put(40,60)\sw\put(60,60)\sw
\put(80,60)\sw\put(30,50)\sw
\put(20,10)\se\put(30,10)\se\put(00,20)\se\put(10,20)\se\put(40,20)\se\put(70,30)\se
\put(00,40)\se\put(50,40)\se\put(20,50)\se\put(00,60)\se\put(30,60)\se\put(60,40)\se
\put(50,60)\se\put(70,60)\se
\put(00,17){\line(0,-1){4}}\put(00,37){\line(0,-1){4}}\put(00,57){\line(0,-1){4}}\put(10,07){\line(0,-1){4}}
\put(10,17){\line(0,-1){4}}\put(10,27){\line(0,-1){4}}\put(10,37){\line(0,-1){4}}\put(10,47){\line(0,-1){4}}
\put(20,07){\line(0,-1){4}}\put(20,17){\line(0,-1){4}}\put(20,27){\line(0,-1){4}}\put(20,47){\line(0,-1){4}}
\put(30,07){\line(0,-1){4}}\put(30,17){\line(0,-1){4}}\put(30,47){\line(0,-1){4}}\put(30,57){\line(0,-1){4}}
\put(40,07){\line(0,-1){4}}\put(40,17){\line(0,-1){4}}\put(40,57){\line(0,-1){4}}\put(50,07){\line(0,-1){4}}
\put(50,17){\line(0,-1){4}}\put(50,37){\line(0,-1){4}}\put(50,57){\line(0,-1){4}}\put(60,07){\line(0,-1){4}}
\put(60,37){\line(0,-1){4}}\put(60,57){\line(0,-1){4}}\put(60,47){\line(0,-1){4}}\put(70,27){\line(0,-1){4}}
\put(70,47){\line(0,-1){4}}\put(70,57){\line(0,-1){4}}\put(80,27){\line(0,-1){4}}\put(80,47){\line(0,-1){4}}
\put(80,57){\line(0,-1){4}}\put(90,47){\line(0,-1){4}}\put(10,57){\line(0,-1){4}}
\put(13,00){\line(1,0){4}}\put(33,00){\line(1,0){4}}\put(53,00){\line(1,0){4}}\put(03,10){\line(1,0){4}}
\put(13,10){\line(1,0){4}}\put(23,10){\line(1,0){4}}\put(33,10){\line(1,0){4}}\put(43,10){\line(1,0){4}}
\put(53,10){\line(1,0){4}}\put(03,20){\line(1,0){4}}\put(13,20){\line(1,0){4}}\put(23,20){\line(1,0){4}}
\put(43,20){\line(1,0){4}}\put(03,30){\line(1,0){4}}\put(13,30){\line(1,0){4}}\put(53,30){\line(1,0){4}}
\put(53,40){\line(1,0){4}}\put(63,40){\line(1,0){4}}\put(83,40){\line(1,0){4}}\put(03,50){\line(1,0){4}}
\put(03,40){\line(1,0){4}}\put(23,40){\line(1,0){4}}\put(73,30){\line(1,0){4}}\put(73,20){\line(1,0){4}}
\put(23,50){\line(1,0){4}}\put(33,50){\line(1,0){4}}\put(53,50){\line(1,0){4}}\put(63,50){\line(1,0){4}}
\put(73,50){\line(1,0){4}}\put(83,50){\line(1,0){4}}\put(03,60){\line(1,0){4}}\put(33,60){\line(1,0){4}}
\put(53,60){\line(1,0){4}}\put(73,60){\line(1,0){4}}
\put(09,68){$a$}\put(98,-2){$b$}\put(54,23){$z$}
\put(10,65){\circle*{2}}
\put(60,20){\circle{6}}
\put(95,00){\circle*{2}}
\end{picture}}
}
\caption{Rearrangement at a point $z$: we only change connections inside
 a small circle marking $z$.
Either interface does not pass through $z$ in both configurations,
or it passes in a way similar to the pictured above.
On the left the interface (in bold) passes through $z$ twice,
on the right (after the rearrangement) it passes once, but a new loop through $z$
appears (also in bold).
The loops not passing through $z$ remain the same,
so the weights of configurations
differ by a factor of $\sqrt{q}$ because of 
the additional loop on the right.
To get some linear relation on values of $F$,
it is enough to check that any pair of such configurations
makes equal contributions to two sides of the relation.}
\label{fig:perestroika}
\end{figure}

The easiest rearrangement involves redirecting curves passing through $z$,
see Figure~\ref{fig:perestroika},
and we have a good control over relative weights of configurations
whenever they are defined through windings.
Counting how much a pair of configurations
contributes to values of $F$ at neighbors of $z$, we
get some relations. 
Moreover, a careful analysis shows that
the maximal number of relations is attained with the complex weight (\ref{eq:fdef}).

\subsection{Ising model}

We finish with a sketch of our proof 
for the random cluster representation
of the Ising model (i.e. $q=2$) on the square lattice $\mesh\ZZ^2$
at the critical temperature.
As before consider loop representation
in a simply connected domain $\Omega$ with two boundary points $a$ and $b$
and Dobrushin boundary conditions.

Consider function $F=F_\epsilon(\Omega,a,b,z)$
given by (\ref{eq:fdef})
which is the expectation that interface 
from $a$ to $b$ passes through a vertex $z$
taken with appropriate unit complex weight.
Note that for Ising $q=2$, so $k=1/8$
and the weight 
is  Fermionic 
(which of course was expected):
a passage in the same direction but with a $2\pi$ twist has a relative
weight $-1$, whereas a passage in the opposite direction with a
counterclockwise $\pi$ twist has a relative weight $-i$.

As discussed $F$ automatically has the martingale property
when we draw $\gamma$ starting from $a$,
so only conformal invariance in the limit has to be checked.

Color lattice vertices in chessboard fashion,
and to each edge $e$ prescribe orientation such that it points
from a black vertex to a white one,
turning it into a vector, or equivalently a complex number $e$.
Denote by $\ell(e)$ the line passing through the origin and
$\sqrt{\bar e}$ -- the square root of the complex conjugate to $e$
(the choice of the square root is not important).
Careful analysis of the rearrangement
on Figure~\ref{fig:perestroika}
shows that $F$ satisfies the following relation:
for every edge $e\in\Omega$ orthogonal projections of the values of $F$
at its endpoints on the line $\ell(e)$ coincide.
We denote this common projection by $F(e)$
as it would also be given
by the same formula (\ref{eq:fdef})
with $z$ taken on the edge $e$
(to be exact one has to divide by $2\cos(\pi/8)$
to arrive at the same normalization).

It turns out to be a form of \emph{discrete analyticity},
and implies (but does not follow from)
the  common definition.
The latter asks for the discrete version
of the Cauchy-Riemann equations $\partial_{i\alpha}F=i\partial_{\alpha}F$
to be satisfied.
Namely for every lattice square the values of $F$ at four corners 
(denoted $u,v,w,z$ in the counter-clockwise direction)
should obey
$$F(z)-F(v)=i\br{F(w)-F(u)}~.$$

\begin{rem}
In the complex plane {\em holomorphic} (i.e. having a complex derivative)
and {\em analytic} (i.e. admitting a power series expansion) functions are the same,
so the terms are often interchanged.
Though the term {\em discrete analytic} is in wide use,
in discrete setting there are no power expansions, so
it would be more appropriate to speak of
{\em discrete holomorphic} (or \emph{discrete regular}) functions.
\end{rem}

As discussed above,
 $F$ solves a discrete version
of the Riemann Boundary Value Problem (\ref{eq:rbvp})
with $\alpha=1-4k=1/2$,
which was solved in the continuum case
by $\sqrt{\Phi'}$.
It remains to show that as the lattice step
goes to zero,
properly normalized $F$ converges to the latter.

A logical thing to do is  to integrate $F^2$ to retrieve $\Phi$.
Unfortunately, the square of a discrete analytic function
is no longer discrete analytic and so cannot be integrated.
However it turns out that there is a unique
function $H=\IM \int F^2 dz$,
which is defined on the dual lattice by
\begin{equation}\label{eq:h}
H(b)-H(w)~=~|F(e)|^2~,
\end{equation}
where edge $e$ separates the centers of two adjacent squares,
black $b$ and white $w$.

After writing (\ref{eq:h}), one checks that
\begin{enumerate}
\item \label{p1} $H$ is well defined and unique up to an additive constant,
\item \label{p2}  $H$ restricted to white (black) squares
is super (sub) harmonic,
\item  \label{p3} $H=1$ on (counterclockwise) boundary arc $ba$ and
$H=0$ on (counterclockwise) boundary arc $ab$,
\item  \label{p4} The (local) difference between $H$ restricted to white and to black squares tends uniformly to zero.
\end{enumerate}

The properties \ref{p1}, \ref{p2} are consequences of discrete analyticity:
\ref{p1} a rather direct one, while \ref{p2}
follows from the identity
$$
\Delta H(u)~=~\pm\abs{F(x)-F(y)}^2~,
$$
where $u$ is a center of white (black) square
with two opposite corner vertices $x$ and $y$
(particular choice is unimportant).
Definition of $F$ implies the property \ref{p3}.
The property \ref{p4} easily follows from a priori estimates
(namely Kaufman-Onsager-Yang results \cite{Kaufman-Onsager,Yang}).
In principle it should also directly follow from the discrete analyticity 
of $F$ and the property \ref{p3}.

We immediately infer that
$H$ converges to $\IM \Phi$, and
after differentiating and taking a square root
we obtain the following
\begin{prop}
Suppose that the lattice mesh $\epsilon_j$ goes to zero
and a lattice domain $\Omega_j$ 
with boundary points $a_j$, $b_j$ 
converges (in a weak sense, e.g. in Carath\'eodory metric) to
a domain $\Omega$ with boundary points $a$, $b$
as $j\to\infty$.
Then away from the boundary there is a uniform convergence:
$$\frac1{\sqrt{\epsilon_j}}\,{F(\Omega_j,a_j,b_j,z)}~\unif~ \sqrt{\Phi'(\Omega,a,b,z)}~$$
\end{prop}
Since by Proposition~\ref{prop:mart}
the function on the right is
a martingale for \sle{16/3},
convergence of the interface
to the Schramm-L\"owner Evolution with $\kappa=16/3$
follows.

\section{Conclusion}

At the moment the approach discussed above works only for a 
(finite) number
of models.
Another notable case when it works
is the usual spin representation of the Ising model
at critical temperature on the square lattice,
pictured on Figure~\ref{fig:ising}, 
where considering a similar observable
(partition function with $+1$ monodromy, cf. (\ref{eq:fdef}))
leads to the martingale (\ref{eq:psi}) 
and to Schramm-Loewner Evolution with $\kappa=3$.
Interestingly,  exactly the same
definition of discrete analyticity
arises.

Analogously, examination of partition function with $+1$ monodromy at $z$
for hexagonal loop models 
(for all values of $n$ at criticality)
suggests its convergence
to conformal martingale (\ref{eq:psi}).
These considerations lead to a new explanation of the
Nienhuis' Conjecture~\ref{conj:nienhuis} for the critical value of $x$.
In this case we firmly believe that our method works all the way for $n=1$
constructing conformally invariant scaling limits for the ${O}(1)$ model,
but convergence estimates still have to be verified.

Two parallel methods, with observables related
to (\ref{eq:phi}) and (\ref{eq:psi}), seem
specially adapted to the square lattice and the hexagonal lattice
correspondingly.
However, the main arguments work for a large family of
four- and trivalent graphs correspondingly.
So we advance towards establishing the universality conjectures.

Though only for a few models the conformal invariance was proved,
the only essential missing step for the remaining ones
is discrete analyticity, and it can be attacked in a great number of ways.

So from our point of view, the perspectives
for establishing conformal invariance
of classical 2D lattice models
are quite encouraging.
Moreover, we can start discussing reasons for universality,
and try to construct the full loop ensemble
starting from the discrete picture.
The approach discussed above is rigorous, but
what makes it (and the whole \slee~subject)
even more interesting is that while borrowing some intuition
from physics, it gives a new way to approach
these phenomena.

\subsection*{Acknowledgments}
Much of the work was completed while the author was a Royal Swedish Academy of Sciences Research Fellow 
supported by a grant from the Knut and Alice Wallenberg Foundation.
The author also gratefully acknowledges support of the Swiss National Science Foundation.

Existence of a discrete analytic function in the Ising spin model
which has potential to imply convergence of interfaces to \sle3
was first noticed by Rick Kenyon and the author based on 
the dimer techniques applied to the Fisher lattice.
However at the moment the Riemann Boundary Value Problem seemed beyond reach.
John Cardy independently observed that (the classical version) of discrete analyticity 
holds for the function (\ref{eq:fdef}) restricted to edges.

I would like to thank Lennart Carleson for introducing me to this area, 
as well as for constant encouragement and advice. 
Most of what I know about lattice models was learnt from others, and I am especially grateful to Michael Aizenman,
John Cardy and Rick Kenyon
for numerous inspiring conversations 
on the subject.
Much of the \slee~ considerations discussed above are due
to Greg Lawler, Oded Schramm and Wendelin Werner.
Finally I wish to thank Dmitri Beliaev, Ilia Binder and Geoffrey Grimmett
for helpful comments on this note.


\frenchspacing
\begin{thebibliography}{35}

\bibitem{Aizenman-statphys}
M. Aizenman,
\newblock The geometry of critical percolation and conformal invariance.
\newblock In \emph{ STATPHYS 19 (Xiamen, 1995)}.  World Sci.
  Publishing, River Edge, NJ, 1996, 104--120.

\bibitem{Aizenman-Burchard}
\newblock M. Aizenman, A. Burchard,
\newblock H\"older regularity and dimension bounds for random curves,
\emph{ Duke Math. J.} \textbf{ 99} (1999), 419--453.

\bibitem{ABNW}
{\newblock M.~Aizenman, A.~Burchard, C.~M.~Newman, D.~B.~Wilson, 
\newblock Scaling limits for minimal and random spanning trees in two dimensions,
\newblock \emph{Random Structures Algorithms} \textbf{15} (1999), 319--367.}

\bibitem{A-book} {
\newblock I. A. Aleksandrov, 
\newblock \emph{ Parametricheskie prodolzheniya v teorii odnolistnykh funktsii}
(Russian).
{[Parametric continuations in the theory of univalent functions]}
\newblock Izdat. ``Nauka'', Moscow, 1976. }

\bibitem{Bauer-Bernard}{
\newblock M.~Bauer, D.~Bernard,
\newblock 2D growth processes: SLE and Loewner chains.
\newblock 2006, \mbox{arXiv:math-ph/0602049}.}

\bibitem{Baxter-book}
R. J. Baxter, 
\newblock \emph{ Exactly solved models in statistical mechanics}.
\newblock Academic Press, London, 1982.

\bibitem{Beffara}
V.~Beffara,
The dimension of the SLE curves.
2002, \mbox{arXiv:math.Pr/0211322}.

\bibitem{Camia-Newman}
F.~Camia, C.~M. Newman, 
{The full scaling limit of two-dimensional critical percolation}. 
2005, \mbox{arXiv:math.Pr/0504036}.

\bibitem{Cardy-92}
{J.~L. Cardy, 
\newblock Critical percolation in finite geometries,
\newblock \emph{ J. Phys. A} \textbf{ 25} (1992), L201--L206.}

\bibitem{Cardy-expo}
{J. Cardy,
\newblock SLE for theoretical physicists,
\newblock \emph{Ann. Physics} \textbf{318} (2005), 81--118.}
\bibitem{Grimmett-book99}
G. Grimmett,
\newblock \emph{ Percolation}.
\newblock Grundlehren der Mathematischen Wissenschaften 321,
\newblock Springer-Verlag, Berlin, second edition, 1999.

\bibitem{Grimmett-book06}
G. Grimmett,
\newblock \emph{ The Random-Cluster Model}.
\newblock Grundlehren der Mathematischen Wissenschaften 333,
\newblock Springer-Verlag, Berlin, 2006.

\bibitem{Kakutani-44}
Sh. Kakutani,
\newblock Two-dimensional {B}rownian motion and harmonic functions,
\newblock \emph{ Proc. Imp. Acad. Tokyo} \textbf{ 20} (1944), 706--714.

\bibitem {Kaufman-Onsager}
{B.~Kaufman, L.~Onsager,
\newblock {C}rystal statistics. {IV}. {L}ong-range order in a binary crystal,
unpublished (1950).}
 
\bibitem{Nienhuis-sle}
{W. Kager, B. Nienhuis},
{A Guide to Stochastic L\"owner Evolution and its Applications},
\emph{ J. of Statistical Physics}, \textbf{115} (2004), 1149--1229.

\bibitem{Kenyon1}
R. Kenyon, 
{Conformal invariance of domino tiling},
\emph{Ann. Probab.} \textbf{28} (2000), 759--795.

\bibitem{Kenyon2}
R. Kenyon,
Dominos and the {G}aussian free field,
\emph{Ann. Probab.} \textbf{29} (2001), 1128--1137.

\bibitem {Kesten-book}
{H. Kesten}, 
\emph{ Percolation Theory for Mathematicians}. 
Birkh\"auser,  Boston, 1982.

\bibitem {Kesten-scaling}
{H. Kesten},
Scaling relations for 2D-percolation, 
\emph{ Comm. Math. Phys.} \textbf{ 109} (1987), 109--156.

\bibitem{Langlands-bams}
R. Langlands, Ph. Pouliot, Y. Saint-Aubin,
\newblock Conformal invariance in two-dimensional percolation,
\newblock \emph{ Bull. Amer. Math. Soc. (N.S.)} \textbf{ 30} (1994), 1--61.

\bibitem{Langlands-ising}
R. P. Langlands, M.-A. Lewis, Y.~Saint-Aubin, 
Universality and conformal invariance for the Ising model in domains with boundary,
\emph{J. Statist. Phys.} \textbf{98} (2000), 131--244.

\bibitem{Lawler-book} 
G. F. Lawler, 
\emph{Conformally Invariant Processes in the Plane}.
Amer. Math. Soc., Providence, RI, 2005.
\bibitem{Lawler-Schramm-Werner-ust}
{G.~F.~Lawler, O.~ Schramm, W.~Werner},
\newblock {Conformal invariance of planar loop-erased random walks and uniform spanning trees},
\newblock \emph{Ann. Probab.} \textbf{32} (2004), 939--995.

\bibitem{Levy-book}
P. L\'evy, 
\emph{Processus Stochastiques et Mouvement Brownien. 
Suivi d'une Note de M. Lo`eve.}
Gauthier-Villars, Paris, 1948.

\bibitem{Loewner}
{K.~L\"owner,}
\newblock Untersuchungen \"uber schlichte konforme Abbildung des Einheitskreises, I.
\newblock \emph{ Math. Ann. } \textbf{ 89},  103--121 (1923).


\bibitem {Madras-Slade}
N. Madras, G. Slade,
\emph{The Self-Avoiding Walk}.
Birkh\"auser,  Boston, 1993.

\bibitem {Mccoy-Wu-book}
{B.~M.~McCoy, T.~T.~Wu}, 
\newblock \emph{ The two-dimensional Ising model}. 
\newblock Harvard University Press, Cambridge, Massachusetts, 1973.

\bibitem{N1}
{B. Nienhuis,
Exact critical point and critical exponents of {${\mathrm O}(n)$} models in two dimensions,
\emph{Phys. Rev. Lett.} \textbf{49} (1982), 1062--1065.}
\bibitem {N2}
{B. Nienhuis,
Coulomb gas description of 2-D critical behaviour,
J. Stat. Phys. \textbf{ 34}  (1984), 731-761.}

\bibitem{Rohde-Schramm}
{S. Rohde, O. Schramm},
{Basic properties of SLE},
\emph{ Ann. Math.} \textbf{161}(2005), 883--924.

\bibitem{rsw1}
L. Russo, 
A note on percolation, 
\emph{Z. Wahrscheinlichkeitstheorie und Verw. Gebiete}
\textbf{43} (1978), 39-–48.

\bibitem{Schramm-lerw}
O. Schramm,
\newblock Scaling limits of loop-erased random walks and uniform spanning trees,
\newblock \emph{ Israel J. Math.} \textbf{ 118}, 221--288 (2000).

\bibitem {Schramm-formula}
{O. Schramm},
A percolation formula,
\emph{ Elect. Comm. Probab.}, 
\textbf{ 6} (2001) 115--120.

\bibitem {Schramm-icm}
{O. Schramm},
Conformally invariant scaling limits,
an overview and a collection of problems.
In \emph{Proceedings of the ICM 2006, Madrid}.
2006, to appear.

\bibitem{Schramm-Sheffield}
O. Schramm, S. Sheffield, 
Harmonic explorer and its convergence to ${\rm SLE}\sb 4$,
\emph{Ann. Probab.} \textbf{33} (2005), 2127--2148.

\bibitem{rsw2}
P. D. Seymour and D. J. A. Welsh, 
Percolation probabilities on the square lattice.
Advances in graph theory (Cambridge
Combinatorial Conf., Trinity College, Cambridge, 1977),
\emph{Ann. Discrete Math.} \textbf{3} (1978), 227–-245.

\bibitem {Smirnov-cras}
{S. Smirnov},
Critical percolation in the plane: Conformal invariance, Cardy's
formula, scaling limits, 
\emph{ C. R. Acad. Sci. Paris} \textbf{ 333}, 239--244 (2001).

\bibitem {Smirnov}
{S. Smirnov},
Critical percolation in the plane.
I. Conformal invariance and Cardy's formula.
II. Continuum scaling limit.
\emph{Preprint}, 2001.

\bibitem {smirnov-fk1}
{S. Smirnov},
Conformal invariance in random cluster models. 
I. Holomorphic fermions in the Ising model.
\emph{Preprint}, 2007.

\bibitem{smirnov-fk2}
Stanislav Smirnov.
Conformal invariance in random cluster models. {II}. {S}caling limit of the interface.
\emph{Preprint}, 2007.

\bibitem{smirnov-fk3}
Stanislav Smirnov.
Conformal invariance in random cluster models. {III}. {F}ull scaling  limit.
\emph{In preparation}, 2007.

\bibitem{smirnov-is}
Stanislav Smirnov.
Conformal invariance in 2D Ising model.
\emph{In preparation}, 2007.

\bibitem{Smirnov-Werner}
S. Smirnov, W. Werner,
 Critical exponents for two-dimensional percolation,
\emph{ Math. Res. Lett.} \textbf{ 8}  (2001), 729-744.

\bibitem{Werner-stflour}
W. Werner,
Random planar curves and Schramm-L\"owner Evolutions,
Lecture notes from the 2002 Saint-Flour summer school.
In \emph{Lectures on probability theory and statistics}, 107--195, Lecture Notes in Math., 1840, Springer, Berlin, 2004.

\bibitem{Werner-loops}{
\newblock W. Werner,
\newblock Some recent aspects of random conformally invariant systems.
\newblock 2005, \mbox{arXiv:math.PR/0511268}.}

\bibitem{Wiener}
N.~Wiener,
Differential space,
\emph{J. Math. Phys.} \textbf{ 58} (1923) 131--174.

\bibitem {Yang}
{C.~N.~Yang,
The spontaneous magnetization of a two-dimensional {I}sing model,
\emph{Physical Rev. (2)} \textbf{85} (1952), 808--816.}

\end{thebibliography}
\end{document}